\documentclass[manuscript,revised]{geophysics}
\usepackage{latexsym,amsmath,amssymb,amsthm,systeme}
\usepackage[toc]{appendix}
\renewcommand{\figdir}{Fig}

\usepackage{setspace}
\usepackage{appendix}

\begin{document}

\title{Seismic Inversion by Hybrid Machine Learning}
\renewcommand{\thefootnote}{\fnsymbol{footnote}}

\ms{Seismic Inversion by Hybrid Machine Learning} 
\address{
\footnotemark[1] Deep Earth Imaging Future Science Platform, CSIRO, Kensington, Australia.
}
\author{Yuqing Chen\footnotemark[1] and Erdinc Saygin\footnotemark[1] }

\footer{Example}
\lefthead{Chen, Saygin \& Schuster}
\righthead{\emph{Seismic Inversion by HML}}

\begin{abstract}

We present a new seismic inversion method which uses the deep learning (DL) features for the subsurface velocity model estimation. The DL feature is a low-dimensional representation of the high-dimensional seismic data, which is automatically generated by a convolutional autoencoder (CAE) and preserved in the latent space. The low-dimensional DL feature contains the key information of the high-dimensional input seismic data. Therefore, instead of directly comparing the waveform differences between the observed and predicted data, such as full waveform inversion (FWI). We measure their DL feature differences in the latent space of a CAE. The advantage of this low-dimensional comparison is that it is less prone to the cycle-skipping problem compared to FWI. The reason is that the automatically generated DL features mainly contain the kinematic information, such as traveltime, of the input seismic data when the latent space dimension is small. However, more dynamic information, such as the waveform variations, can be preserved in the DL feature when the latent space dimension becomes larger. Therefore we propose a multiscale inversion approach that starts with inverting the low-dimensional DL features for the low-wavenumber information of the subsurface velocity model. Then recover its high-wavenumber details through inverting the high-dimensional DL features. However, there is no governing equation that contains both the velocity and DL feature terms in the same equation. Therefore we use the automatic differentiation (AD) to numerically connect the perturbation of DL features to the velocity perturbation. In another word, we connect a deep learning network with the wave-equation inversion by using the AD. We denote this hybrid connection as hybrid machine learning (HML) inversion. Here, the AD replaces the complex math derivations of gradient with a black box so anyone can do HML without having a deep geophysical background.

One concern of the HML method is that it is expensive to solve the wave-equation inversion using the AD. To mitigate this problem, we propose a hybrid implementation approach that uses the AD only through the CAE and compute the velocity gradient using imaging condition. This approach is computational efficient and benefit from the quasi-linear misfit function at the same time. Numerical tests on both synthetic and real data show that the multi-scale HML approach can effectively recover both the low- and high-wavenumber information of the subsurface velocity model.

\end{abstract}

\section{Introduction}
Full waveform inversion (FWI) is a powerful tool for inverting the high-resolution subsurface model by minimizing the waveform differences between the observed and predicted data \citep{lailly1983seismic, tarantola1984inversion, virieux2009overview, simute2016full, perez2019velocity}. However, the conventional FWI assumes its forward modeling operator $\mathbf{L}$ includes all the physics of wave propagation in the real Earth. Moreover, a good initial model is essential for FWI which requires the time-lag between the observed and predicted data should be smaller than half of the period. Otherwise FWI will suffer from the cycle-skipping problem and the failure of these assumptions could lead FWI to converge to a local minimum. To mitigate these problems, an alternative solution is to invert the skeletonized data, such as the traveltime and peak frequency, rather than the whole waveform. The skeletonized data is a simplified form of the original data but still contains its key information. Therefore inverting the skeletonized data is less prone to the local minimum and can successfully recover the low-to-intermediate wavenumber information of the model of interests. \cite{luo1991wavea, luo1991waveb} used the wave-equation solution to invert the first arrival traveltime for the subsurface background velocity model. \cite{dutta2016wave} inverted for the Qp model by minimizing the central/peak frequency differences between the observed and predicted early arrivals. Similarly, \cite{li2017wave} utilized the peak frequency shifts of the surface wave to invert for the Qs model. \cite{li2016wave} and \cite{liu20183d} found the optimal S-velocity model by using the dispersion curves associated with the surface waves. A more comprehensive introduction of the skeletonized inversion can be found in \cite{lu2017tutorial}.

The generation of skeletonized data mentioned above are based on human knowledge and usually require manual picking. For a large dataset, this picking task is labor intensive and time consuming. Here we use a convolutional neural network (CAE) to automatically extract the skeletal information from seismic data and no manual picking is required. The skeletal data, also know as the deep learning (DL) feature, is a low-dimensional representation of the high-dimensional input seismic trace, which contains the key information of the input data and preserved in the latent space of the CAE. When the latent space dimension is small, the DL feature mainly contains the kinematic information of the seismic trace, such as traveltime. However, a high dimensional latent space is capable of preserving both the kinematic and dynamic information of seismic data. In this paper, we invert for the subsurface velocity by measuring the difference of the DL feature between the observed and predicted data in the latent space of a well-trained CAE. We first invert the low-dimensional DL features for the low-wavenumber information of subsurface velcoity model. We then recover its high-wavenumber details by inverting the high-dimensional DL features. However, there is no governing equation contain both DL feature and velocity terms in the same equation. Therefore there is no way to connect the perturbation of the DL feature to the velocity perturbation directly. In other words, one can not directly compute the velocity gradient associated with the DL feature differences misfit function. In this research, instead of using the connective function assumption \citep{chen2020seismic}, we use automatic differentiation (AD) to automatically compute the derivative of the DL feature with respect to the velocity model. Therefore the AD technique can numerically connect the CAE network with wave-equation inversion where no assumptions are made. We denote this hybrid connection technique as hybrid machine learning (HML) inversion.

\begin{figure}[h]
\centering
\includegraphics[width=0.9\columnwidth]{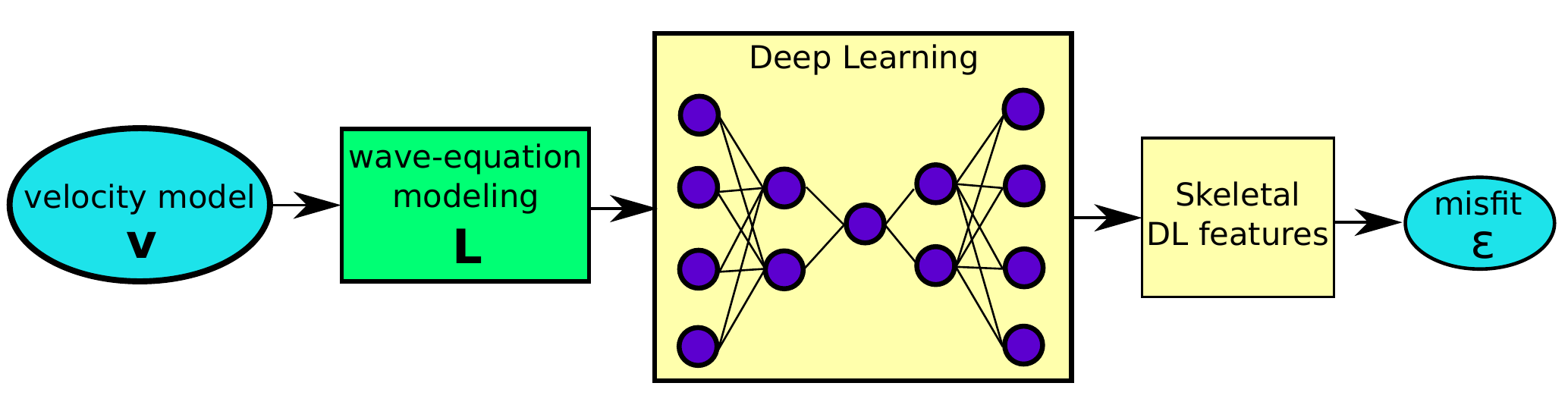}
\caption{The demonstration of the forward progress of HML.}
\label{fig:HML1}
\end{figure}

The AD is a set of techniques to numerically evaluate the derivatives of a function specified by a computer program \citep{schuster2020seismic}. It uses the chain rules to break up the derivative of a complicated composite function into a chain of simple derivatives \citep{schuster2020seismic}. The AD is widely used in deep learning to compute the gradients of the model parameters and bias terms of a DL network. Moreover, the AD has shown its potential in solving the inverse problem. \cite{sambridge2007automatic} showed several examples of using the AD to solve the geophysical inverse problem, such as ray tracing. \cite{hughes2019wave} showed that the wave-equation forward modeling is equivalent to a recurrent neural network (RNN). \cite{sun2020theory} used the AD as an alternative of the imaging condition to compute the FWI gradient. Therefore we use AD as a perfect tool to connect the CAE network with wave-equation inversion, where we only need to program the forward progress (shown in Figure \ref{fig:HML1}) from the velocity model $v$ to the final misfit $\epsilon$. Then the AD can automatically compute the derivative of the misfit to the input velocity model. Here the AD replaces the complex math derivation of $\frac{\partial \epsilon}{\partial v}$ with a black box so anyone can do HML without having a deep geophysical background. Moreover, the CAE network can be replaced by any other networks and the wave-equation can be replaced by other types of Newton equations to solve a variety of problems. No matter what changes have made, the AD can automatically compute the derivative of the misfit with respect to the model of interets.

In HML, a convolutional autoencoder (CAE) is first trained by the seismic traces to learn its low-dimensional DL features that contain the key information of input seismic traces. We then compute the L2 misfit $\epsilon$ of the DL features between the observed and synthetic data in the latent space of the well-trained CAE. Next, the AD computes the velocity gradient $\frac{\partial \epsilon}{\partial v}$ automatically and we use the gradient descent method to update the velocity model. However, one concern of the HML method is that it is computationally expensive to use the AD to solve the wave equation inversion. Because it needs to compute at least $nt \times N$ local derivatives, where $nt$ is the simulation time in time samples and $N$ defines the model size in grid points. For a large 3D model, this computation becomes near impossible. As an alternative, we only use the AD only through the CAE to compute $\frac{\partial \epsilon}{\partial d}$, where $d$ represents the predicted data. We then use the imaging condition to compute the velocity gradient where $\frac{\partial \epsilon}{\partial d}$ is used as the virtual source for constructing the backward-propagated wavefield and then zero-lag cross-correlated with the forward wavefield. This hybrid implementation approach enjoys both the computational efficiency and the quasi-linear property of HML misfit function at the same time, which bring HML the potential for solving the large-scale inversion problems. Numerical tests on both synthetic and real data show that the HML approach can successfully recover the low- and high-wavenumber information of the subsurface velocity model in a multiscale way.


\section{Theory}

\subsection{Convolutional neural network}
\begin{figure}[h]
\centering
\includegraphics[width=1\columnwidth]{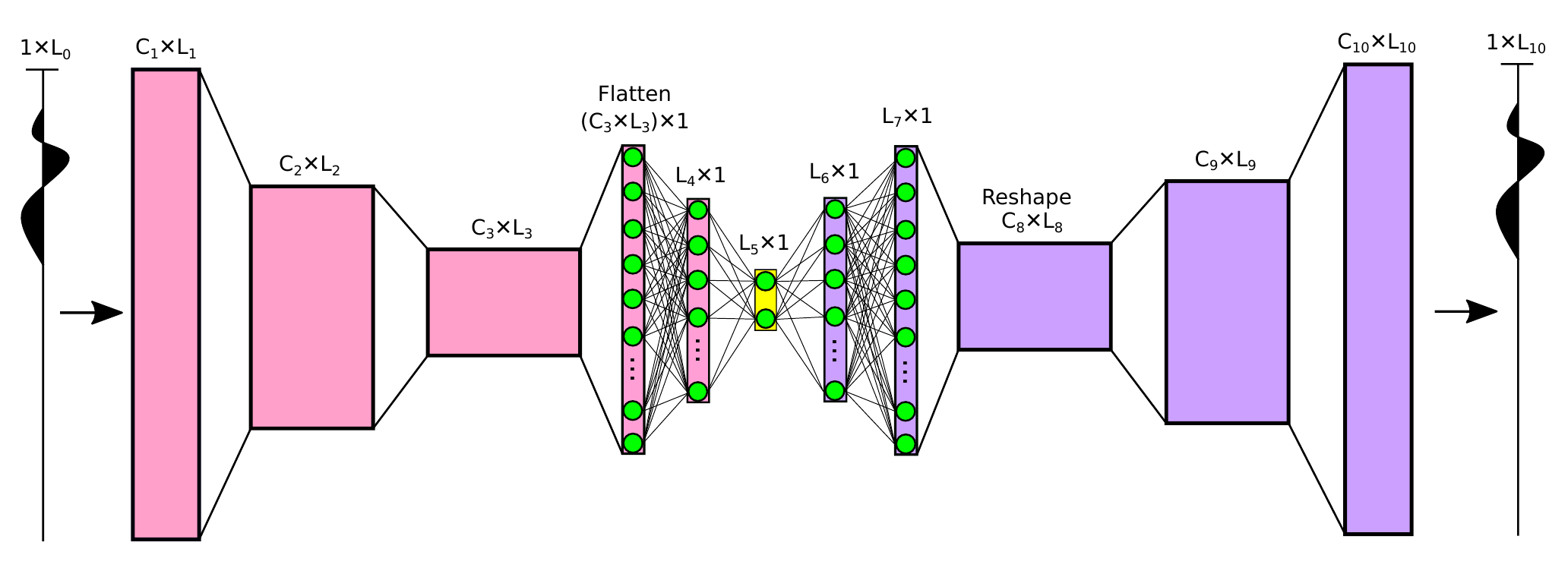}
\caption{An example of a simple function.}
\label{fig:CAE1}
\end{figure}

Convolutional autoencoder (CAE) is an unsupervised neural network that is trained to learn the low-dimensional representation of the high-dimensional input data. An example of a typical 1D CAE architecture is shown in Figure \ref{fig:CAE1}, where the pink, yellow, and purple boxes represent the encoder network, latent space and decoder network, respectively. The encoder network includes three convolutional layers with an increasing number of channels $C$ and decreasing of length $L$. Usually, the "convolution" + activation function + pooling operations exist between each convolutional layer and decide the channel size and length of the next convolutional layer. The data in the last convolutional layer with the size of $C_{3} \times L_{3}$ needs to be flattened to a vector shape with the size of $(C_{3} \times L_{3})\times 1$ to input into the FC layers. There are two FC layers in the encoder network with a decreasing number of neurons in each layer that compresses the high-dimensional input data to the low-dimensional latent space. The yellow box indicates the latent space which preserves the lowest-dimensional DL features which contains the key information of the input data. In this example, the decoder network is the mirror of the encoder network which gradually expands the low-dimensional representation to its original size. The CAE architecture can be iteratively trained by minimizing the L2 misfit between the input and decoded data.  

\subsection{Automatic differentiation}
\begin{figure}[h]
\centering
\includegraphics[width=0.8\columnwidth]{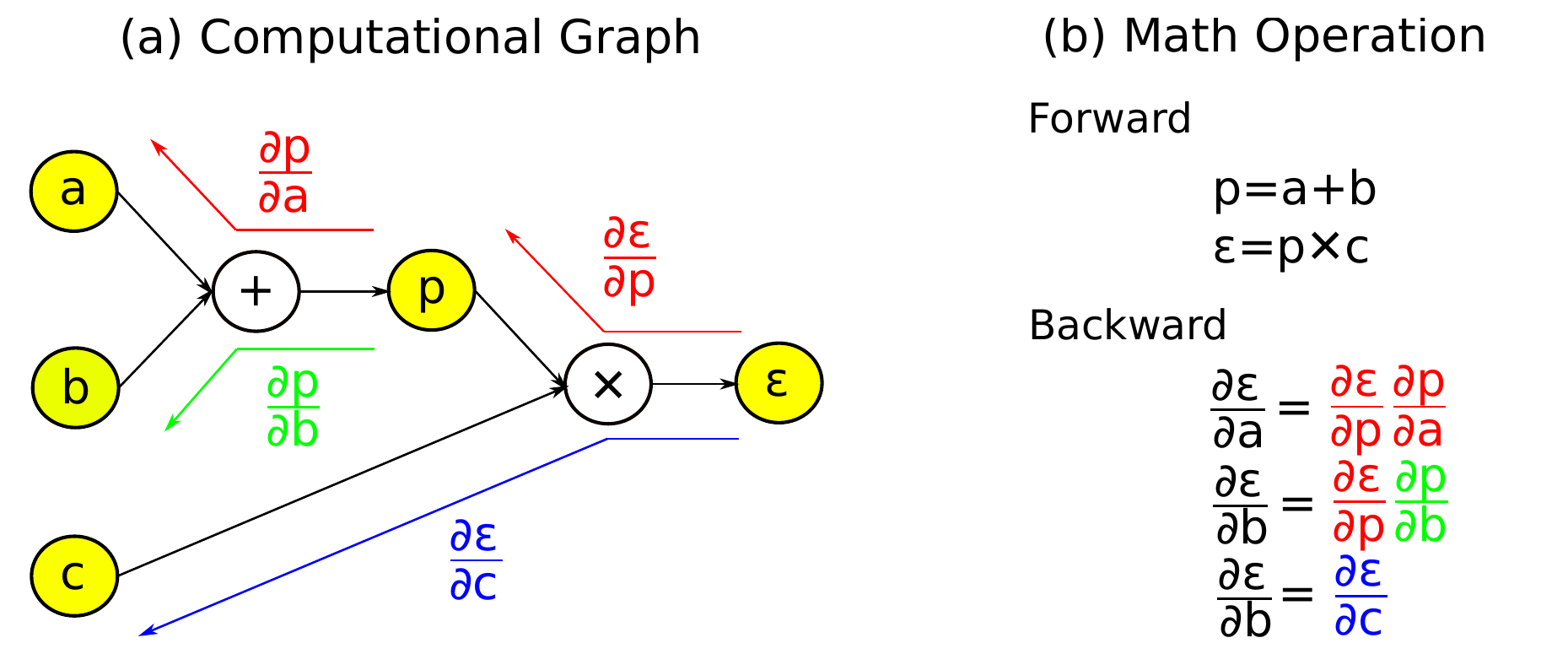}
\caption{(a) The computational graph of the function $\epsilon=(a+b)\times c$ and the (b) math operations of the computational graph. The forward indicates the feedforward operation of the computational graph and the backward indicates the reverse model of the AD, where each local derivative is computed by the AD from the very final misfit $\epsilon$ to the input variables.}
\label{fig:AD1}
\end{figure}

The automatic differentiation (AD) is a technique that numerically estimates the derivative of a function specified by a computer program \citep{schuster2020seismic}. The AD believes that any complicated function is composed by the elementary math operations, such as addition, multiplication, log, exp, etc. Therefore the AD uses the chain rule to break up the derivative of a complicated composite function into a chain of simple derivatives. Figure \ref{fig:AD1} shows an example of computing the derivatives of the function $\epsilon=(a+b)\times c$ using the AD. This function is described by a computational graph in Figure \ref{fig:AD1}a, where the yellow and white nodes indicate the computational and math operations node, respectively. In the forward operation, an intermediate-term $p$ is first generated to represents the result of $a+b$, and then multiplied with $c$ to get the output $\epsilon$. In the backward operation, the AD first computes the derivative of $\epsilon$ to the intermediate variable $p$, then calculates the derivative of the intermediate variable $p$ to each input variable. In general, the AD only computes the local derivative between a pair of the nearby computational node that is directly connected to a math operation node. These computations start from the very final output and way back to the input, this procedure is also denoted as the reverse mode of the AD. Once the AD has computed all the local derivatives, the global derivative, such as $\frac{\partial \epsilon}{\partial a}$, can be acquired by multiplying those local derivatives on a certain computational path. 

\begin{figure}[h]
\centering
\includegraphics[width=0.95\columnwidth]{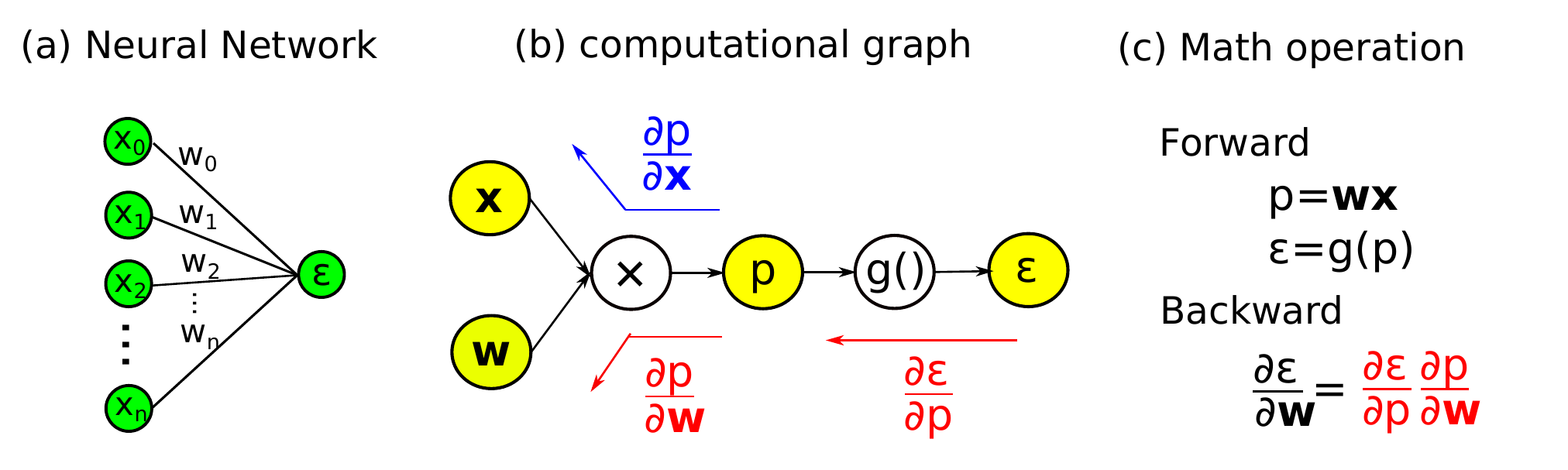}
\caption{A (a) neural network and (b) its computational graph. The (c) forward and backward operation of AD.}
\label{fig:AD2}
\end{figure}

Similarly, a neural network (NN) shown in Figure \ref{fig:AD2}a can be also depicted by a computational graph shown in Figure \ref{fig:AD2}b. Here, $\mathbf{w}$ and $\mathbf{x}$ represent the model parameters of the NN network and the input data, respectively. The forward operation in Figure \ref{fig:AD2}c is very similar to the previous example except the input variables are vectors. Here, $g()$ represent a activations function, such as sigmoid function $\frac{1}{1+e^{-x}}$. To compute $\frac{\partial \epsilon}{\partial \mathbf{w}}$, AD computes each local derivatives from the output back to inputs. And $\frac{\partial \epsilon}{\partial \mathbf{w}}$ can be acquired by multiplying all the local derivatives together along the red path as $\frac{\partial \epsilon}{\partial \mathbf{w}}=\frac{\partial \epsilon}{\partial p}\frac{\partial p}{\partial \mathbf{w}}$. 

\subsection{Hybrid machine learning inversion}
Full waveform inversion (FWI) is a powerful tool in recovering a high-resolution subsurface velocity model by minimizing the waveform differences between the observed and predicted data. However, the FWI misfit function is often characterized by many local minima which could due to many reasons, such as: (1) the wave-equation forward modeling operator $L$ can't take into account all the physics in the real Earth, (2) the initial model is far away from the true model where the time-lag between the observed and predicted data is larger than half of the fundamental period, where FWI suffers from the cycle-skipping problem. To mitigate these problems, instead of computing their waveform differences, we measure their low-dimensional deep learning (DL) feature differences in the latent space of CAE

\begin{equation}
\epsilon = \sum_{s}\sum_{r}\sum_{k} [z_{k}^{obs}(\mathbf{x}_{r},\mathbf{x}_{s})-z_{k}^{pred}(\mathbf{x}_{r},\mathbf{x}_{s})]^{2},
\label{eq:misift1}
\end{equation}
where $z_{k}^{obs}$ and $z_{k}^{pred}$ represents the compressed DL features of the observed and predicted data in the $k_{th}$ latent space dimension. $\mathbf{x}_{s}$ and $\mathbf{x}_{r}$ indicates the locations of source and receiver, respectively. When the latent space dimension is small, the compressed DL feature mainly contains the kinematic information of the seismic data, such as traveltime. Therefore, the HML misfit function in equation \ref{eq:misift1} is less prone to local minima compared to the FWI misfit function. The low-wavenumber information of the subsurface velocity model can be recovered by inverting these low-dimensional DL features. However, more dynamic information such as the waveform variation can be preserved in the DL feature when the latent space becomes larger. As a consequence, the HML method can recover the high-wavenumber information of the subsurface model. Therefore, we propose a multiscale HML inversion approach where we start from inverting the low-dimensional DL features for the low-wavenumber information of the subsurface model. We then recover the high-wavenumber information by inverting the high-dimensional DL features. Similar to FWI, the velocity gradient $\gamma (\mathbf{x})$ can be computed by taking the derivative of misfit $\epsilon$ to the velocity $v$

\begin{equation}
\gamma(\mathbf{x})=-\frac{\partial\epsilon}{\partial v(\mathbf{x})}=-\sum_{s}\sum_{r}\sum_{k} \bigg[ \bigg( \frac{\partial \Delta z_{k}(\mathbf{x}_{r},\mathbf{x}_{s})}{\partial v(\mathbf{x})}\bigg)^{T}\Delta z_{k}(\mathbf{x}_{r},\mathbf{x}_{s})   \bigg].
\label{eq:grad1}
\end{equation}
Because there is no governing equation which contains both the velocity term $v$ and DL features $z$ in the same equation. Therefore there is no way to compute $\frac{\partial z}{\partial v}$ directly. \cite{chen2020seismic} proposed a Newtonian machine learning (NML) inversion which uses a connective function to connect the perturbation of DL feature to the velocity perturbation. However, one problem of the connective function assumption is that, for a multi-dimensional latent space, each latent space dimension is characterized by a gradient and the weighted sum of all these gradients can be used for velocity updates \citep{chen2020multiNML}. Therefore the complexity of NML in both theoretical and computational aspects will increase when the latent space dimension increases. 



\begin{figure}[h]
\centering
\includegraphics[width=0.85\columnwidth]{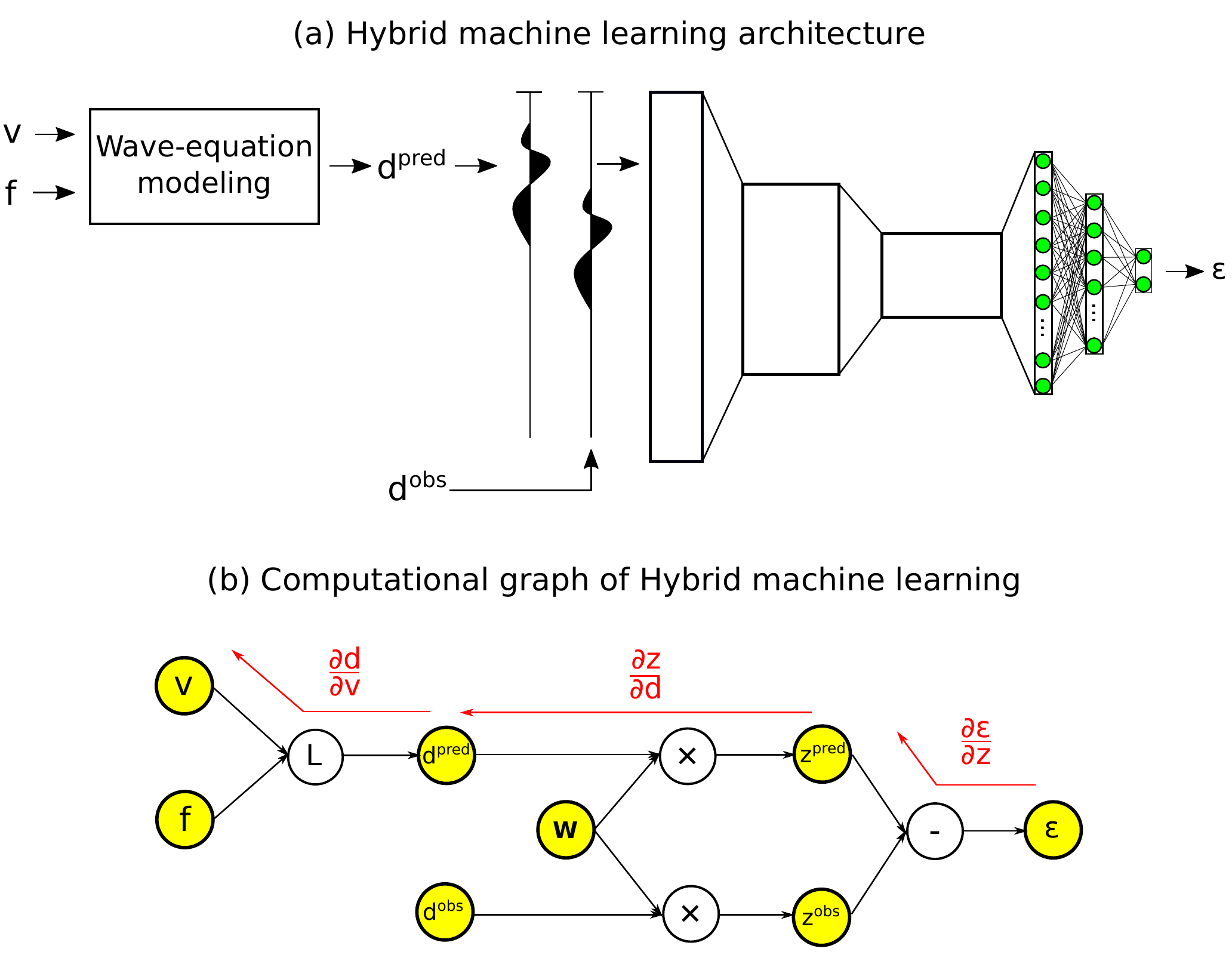}
\caption{The (a) architecture of hybrid machine learning and its (b) simplified version of computational graph.}
\label{fig:AD3}
\end{figure}

\cite{hughes2019wave} and \cite{sun2020theory} showed that the wave-equation modeling is equivalent to the recurrent neural network (RNN) and the FWI gradient can be automatically calculated by the AD. Because CAE training also relies on the AD, therefore the AD is a perfect tool to numerically connect a CAE architecture to the wave-equation inversion. Figure \ref{fig:AD3}a shows the architecture of HML, where we first input a velocity model $v$ and a source wavelet $f$ into a wave-equation modeling module to generate the predicted data $d^{pred}$. We then use the encoder network of a well-trained CAE to compress the observed and synthetic data. Once we get their DL features in the latent space, we compute their L2 misfit using equation \ref{eq:misift1}. This feedforward progress can be described by a simplified computational graph shown in Figure \ref{fig:AD3}b, where $\mathbf{w}$ represents the model parameters of an encoder network from a well-trained CAE. Here, $L$ indicates the wave-equation modeling operation. The symbol $\times$ and $-$ represents the CAE encoding and misfit calculation operation, respectively. All these three operations are composed of elementary math operations such as addition, multiplication, log, and so on. But we do not show their detailed computational graph here otherwise that will be too complicated. Once you have programmed the feedforward progress from the velocity $v$ to the misfit $\epsilon$, the AD can automatically compute each local derivatives, such as $\frac{\partial \epsilon}{\partial z}$, from the very final misfit $\epsilon$ way back to the input velocity model $v$. Therefore, the global derivative $\frac{\partial \epsilon}{\partial v}$, which is the velocity gradient regarding the HML misfit function, can be computed by multiplying all of the local derivatives together which located on the computational path indicated by the red line in Figure \ref{fig:AD3}b. In summary, the AD can automatically compute the velocity gradient once you have programmed the feedforward progress, where no connective function assumption is required and no need to derive the complicated formula of the imaging condition. The AD replaces these complex math derivations with a black box so anyone can do HML without having a deep background in geophysics. Moreover, the CAE network can be replaced by any other deep learning architecture and the wave-equation can be replaced by other Newton equations to solve a variaty of problems. However, no matter what changes are made, the AD can still automatically compute the derivative of the misfit with respect to the model of interets. 

\subsubsection{Hybrid machine learning using a hybrid implementation approach}
\begin{figure}[h]
\centering
\includegraphics[width=0.7\columnwidth]{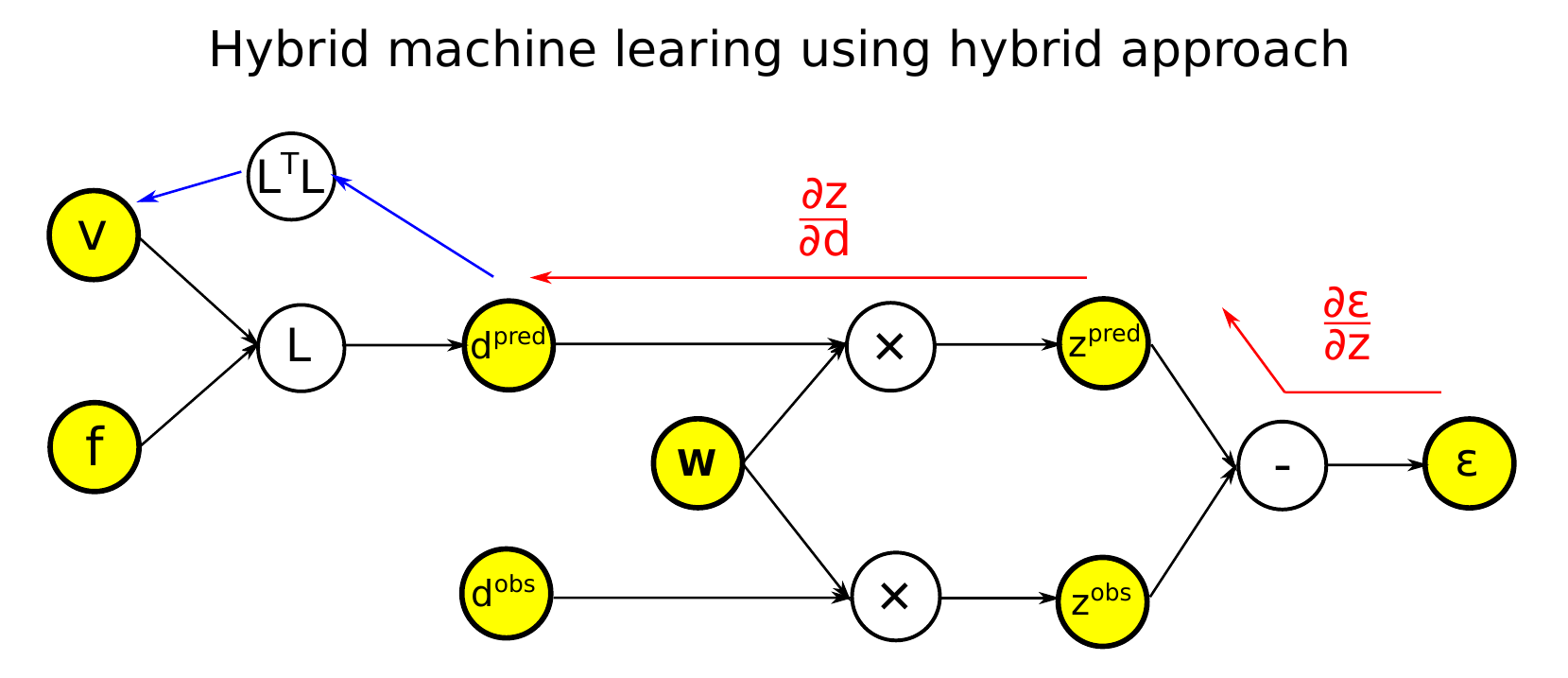}
\caption{The architecture of hybrid machine learning with the hybrid implementation approach.}
\label{fig:AD4}
\end{figure}

Using the AD to solve the wave-equation inversion is computationally expensive. Because it needs to compute at least $nt \times N$ local derivatives, where $nt$ is the simulation time in time samples and $N$ defines the model size in grid points. For a large 3D inversion project, this computation task becomes near impossible. To mitigate this problem, we propose a hybrid implementation approach where we only use the AD through the CAE to compute $\frac{\partial \epsilon}{\partial d}$ and then use the imaging condition to compute the velocity gradient $\frac{\partial \epsilon}{\partial v}$. Here, the AD computed derivative $\frac{\partial \epsilon}{\partial d}$ is used as the virtual source to construct the backward propagated wavefield, which is then zero-lag cross-correlated with the forward wavefield to generate the velocity gradient $\frac{\partial \epsilon}{\partial v}$. Figure \ref{fig:AD4} shows the computational graph of HML using the hybrid implementation approach, which is very similar to Figure \ref{fig:AD3}b. The only difference is that the calculation of $\frac{\partial d}{\partial v}$ is replaced by the wave-equation inversion kernel $L^{T}L$. Therefore the velocity gradient $\frac{\partial \epsilon}{\partial v}$ can be expressed as

\begin{equation}
\frac{\partial \epsilon}{\partial v} = L^{T}L(\frac{\partial z}{\partial d}\frac{\partial \epsilon}{\partial z}). 
\label{eq:connect2}
\end{equation}
Because the computation cost of $\frac{\partial z}{\partial d}$ and $\frac{\partial \epsilon}{\partial z}$ is trivial compared to $L^{T}L$. Therefore the computational efficiency of HML with the hybrid implementation approach is approximately equal to the conventional inversion method, such as FWI. However, HML is less prone to the local minima by inverting the low-dimensional DL features. But also can recover the high-wavenumber details through inverting the higher-dimensional DL feature. This multiscale inversion strategy guarantees that HML with the hybrid implementation approach can efficiently recover a reliable subsurface velocity for both its low- and high-wavenumber information.

\section{Numerical Tests}
In the numerical tests, the HML with the hybrid approach is first tested by two synthetic datasets with the corsswell geometry. We then test this method using a field dataset collected at the Gulf of Aqaba by a surface geometry. In the descriptions below, $la=n$ represents the latent space dimension equal to $n$, where $n$ is a real number.

\subsection{Layered model}
\begin{figure}[h]
\centering
\includegraphics[width=0.6\columnwidth]{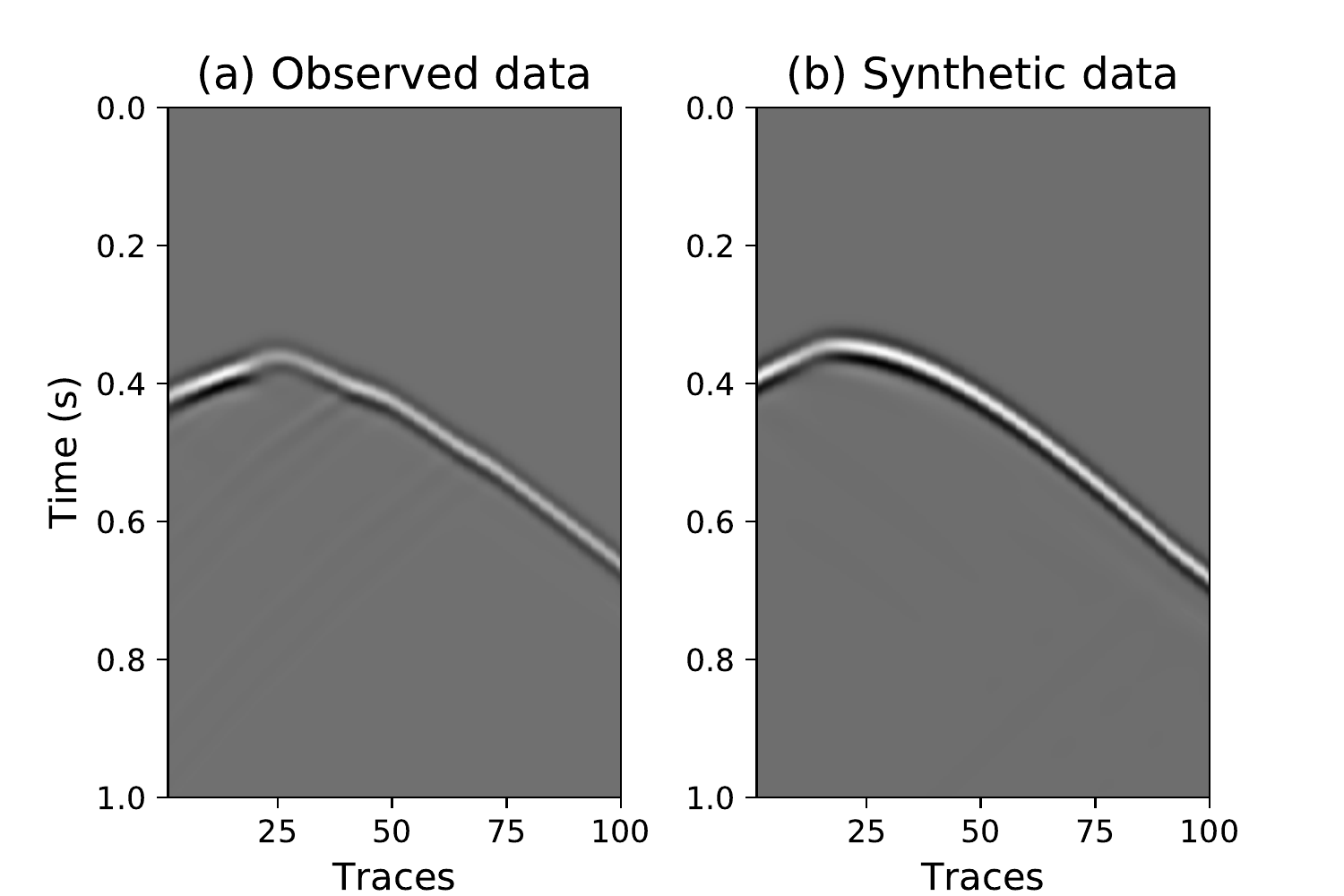}
\caption{One example of the (a) observed and (b) predicted common shot gather.}
\label{fig:L1}
\end{figure}

\begin{figure}[h]
\centering
\includegraphics[width=1\columnwidth]{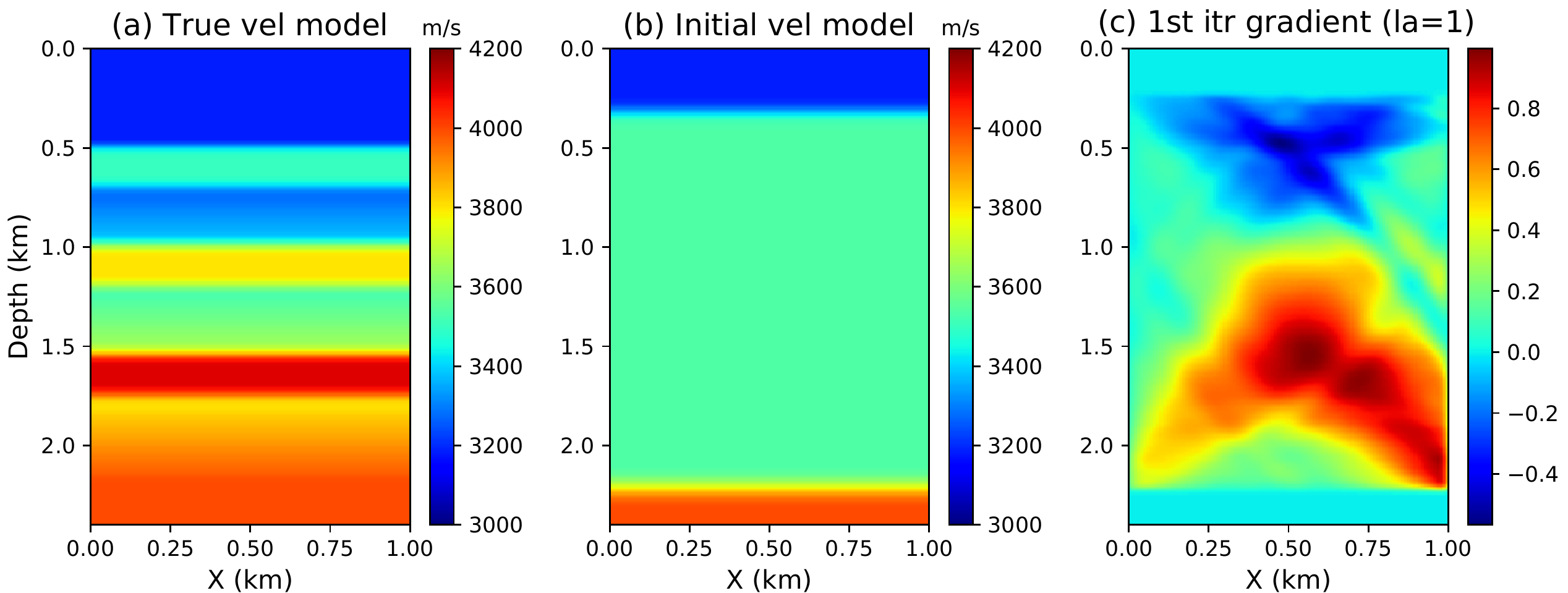}
\caption{The (a) true and (b) initial model. The (c) first iteration gradient of HML.}
\label{fig:L4}
\end{figure}

A layered model with an linear increasing background is used as the true model which is shown in Figure \ref{fig:L4}a. Figure \ref{fig:L4}b shows the initial model where the effective inversion area between z = 0.4 km to z = 2.2 km is set as a homogeneous model with a constant velocity equals to 3535 m/s. 119 acoustic shots are generated by a crosswell acquisition system where the source and receiver well are located at x = 0.01 km and x = 1 km, respectively. These shots are evenly distributed on the source well at an interval of 20 m. Each shot has 239 receivers deployed on the receiver well at an equal spacing of 10 m. A 20 Hz Ricker wavelet is used as the source wavelet. Figures \ref{fig:L1}a and \ref{fig:L1}b show one example of the observed and predicted data, where most of the traces are suffers from the cycle-skipping problem. 

\begin{figure}[h]
\centering
\includegraphics[width=0.65\columnwidth]{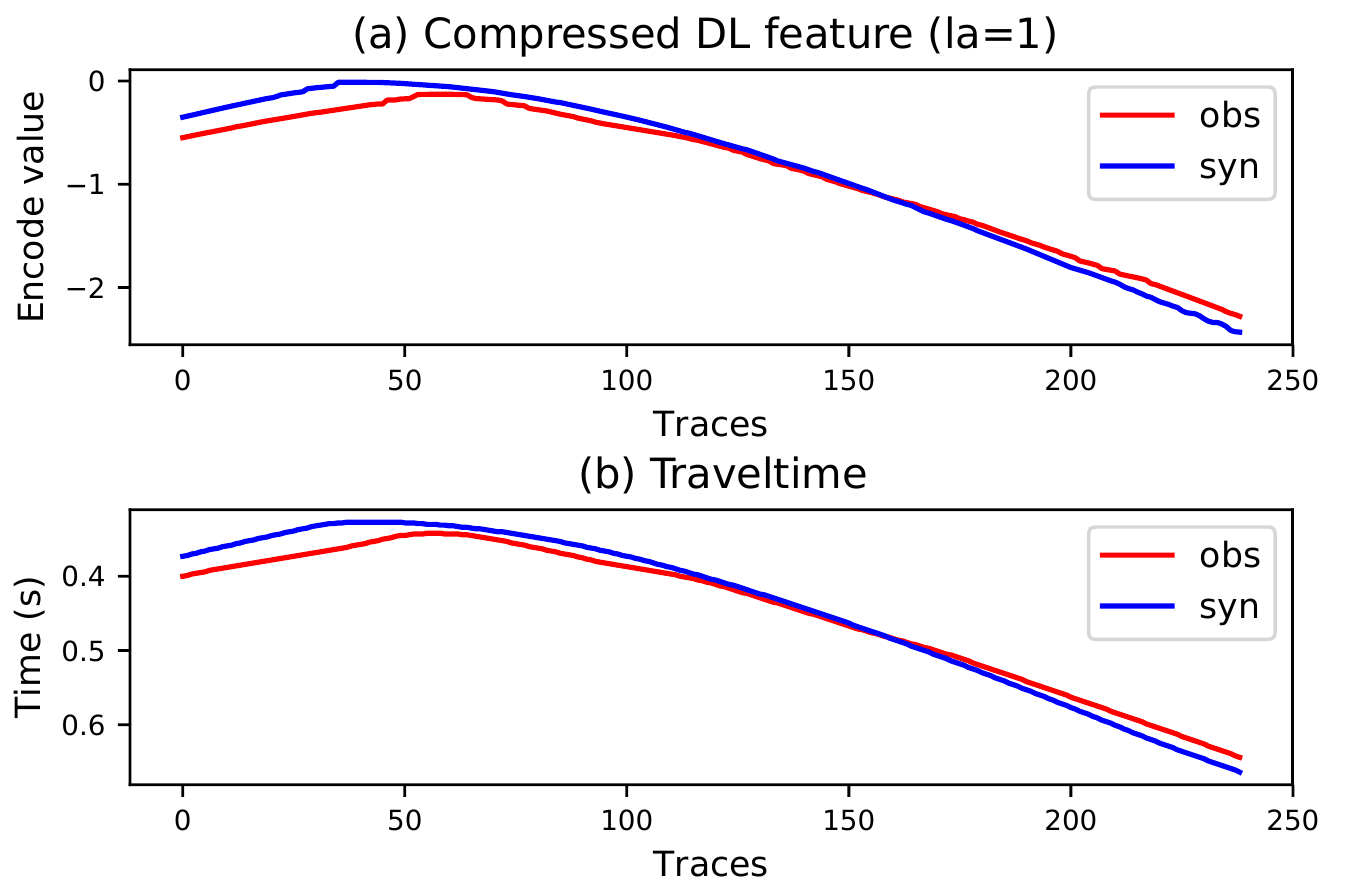}
\caption{The (a) compressed one-dimensional DL features and (b) traveltime.}
\label{fig:L2}
\end{figure}

Before HML inversion, an autoencoder needs to be trained to learn the low-dimensional DL features of the input data. We use the seismic traces from the observed and predict shot gathers as the training data to train an autoencoder with the latent space dimension equals to one. Here, each $nt \times 1$ seismic trace is first compressed to a $1 \times 1$ DL feature by the encoder network and then decoded back to $nt \times 1$ by using the decoder network. Figure \ref{fig:L2}a shows the compressed one-dimensional DL feature of the observed and predicted data shown in Figure \ref{fig:L1}a and \ref{fig:L1}b,  which are represented by the red and blue curves, respectively. The compressed DL features are very similar to the traveltime shown in Figure \ref{fig:L2}b. This similarity demonstrates that the compressed $1\times1$ DL features mainly preserves the kinematic information of the input seismic trace. Therefore the HML misfit function is characterized by less local minima compared to the FWI misfit function. 

\begin{figure}[h]
\centering
\includegraphics[width=0.6\columnwidth]{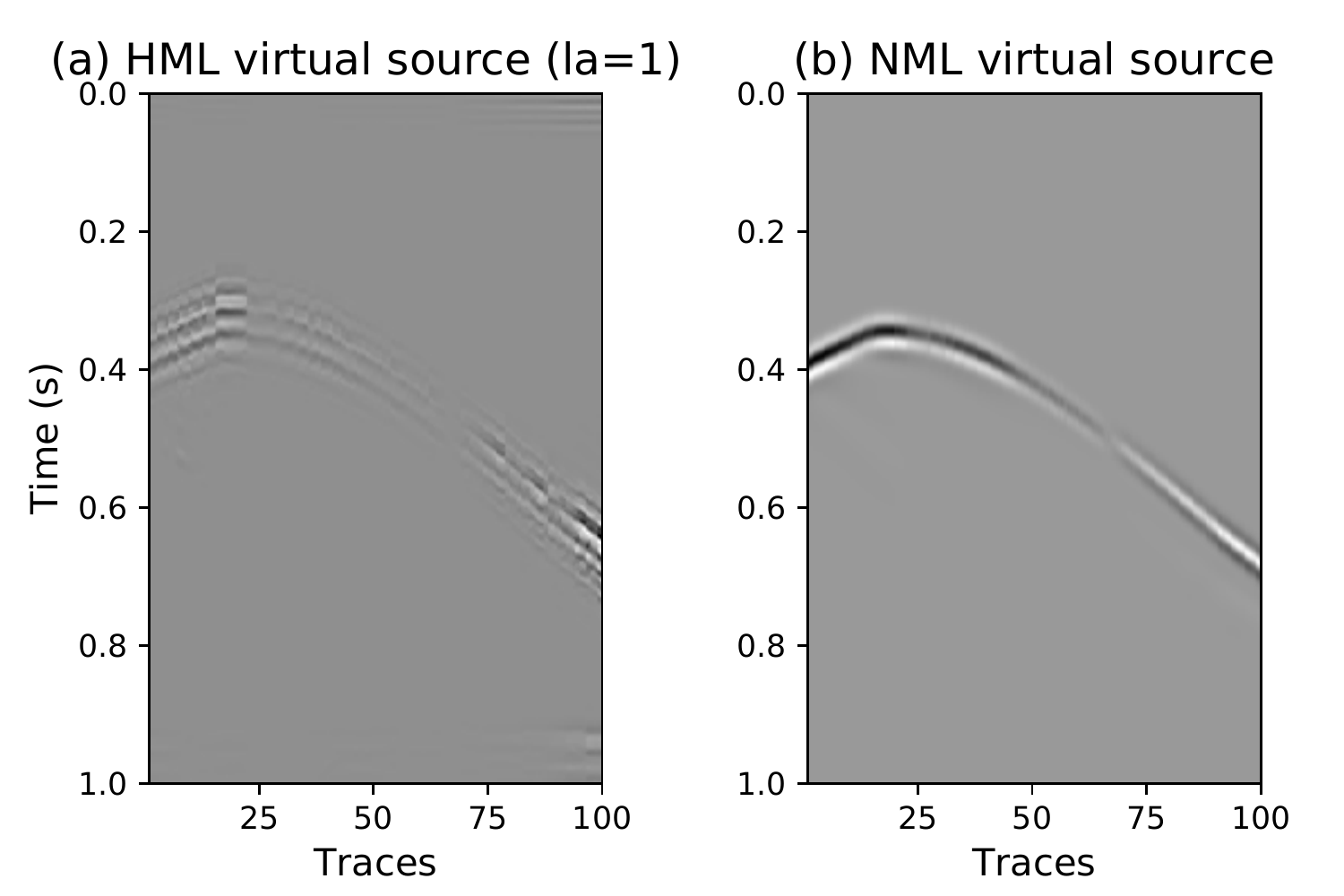}
\caption{The virtual source of (a) HML and (b) NML.}
\label{fig:L3}
\end{figure}

\begin{figure}[h]
\centering
\includegraphics[width=1\columnwidth]{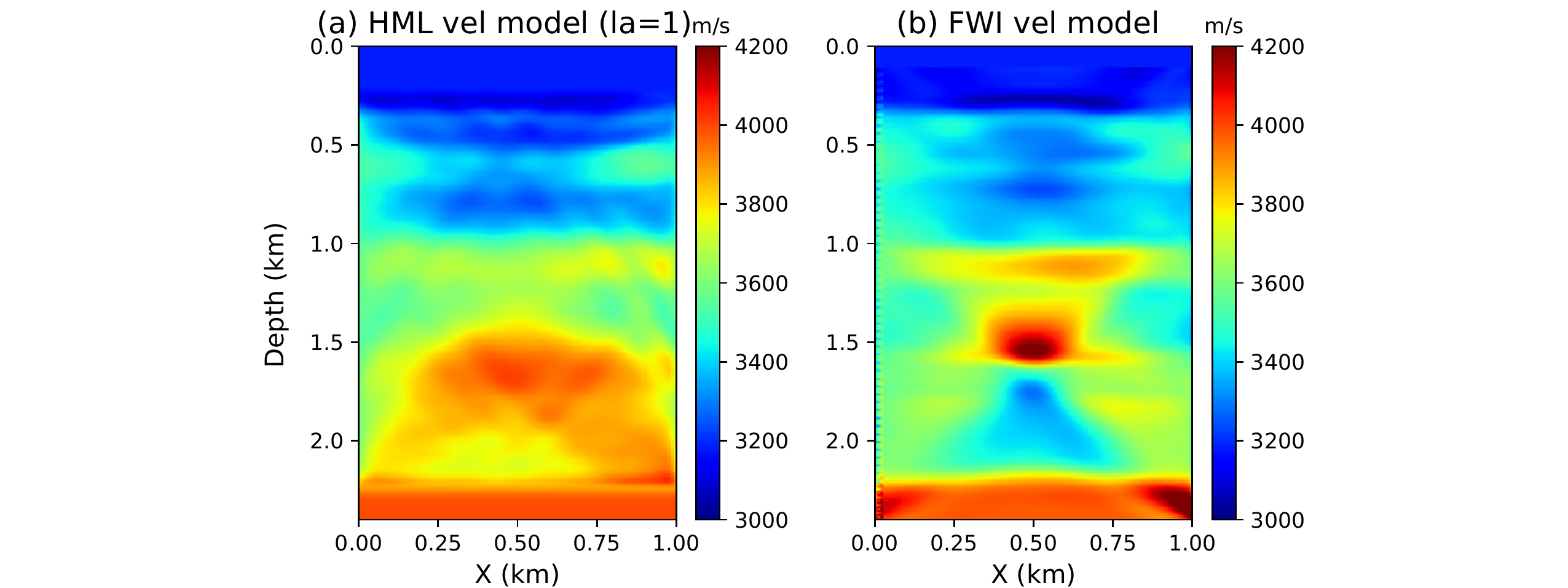}
\caption{The (a) HML and FWI inverted velocity model.}
\label{fig:L5}
\end{figure}

\begin{figure}[h]
\centering
\includegraphics[width=0.75\columnwidth]{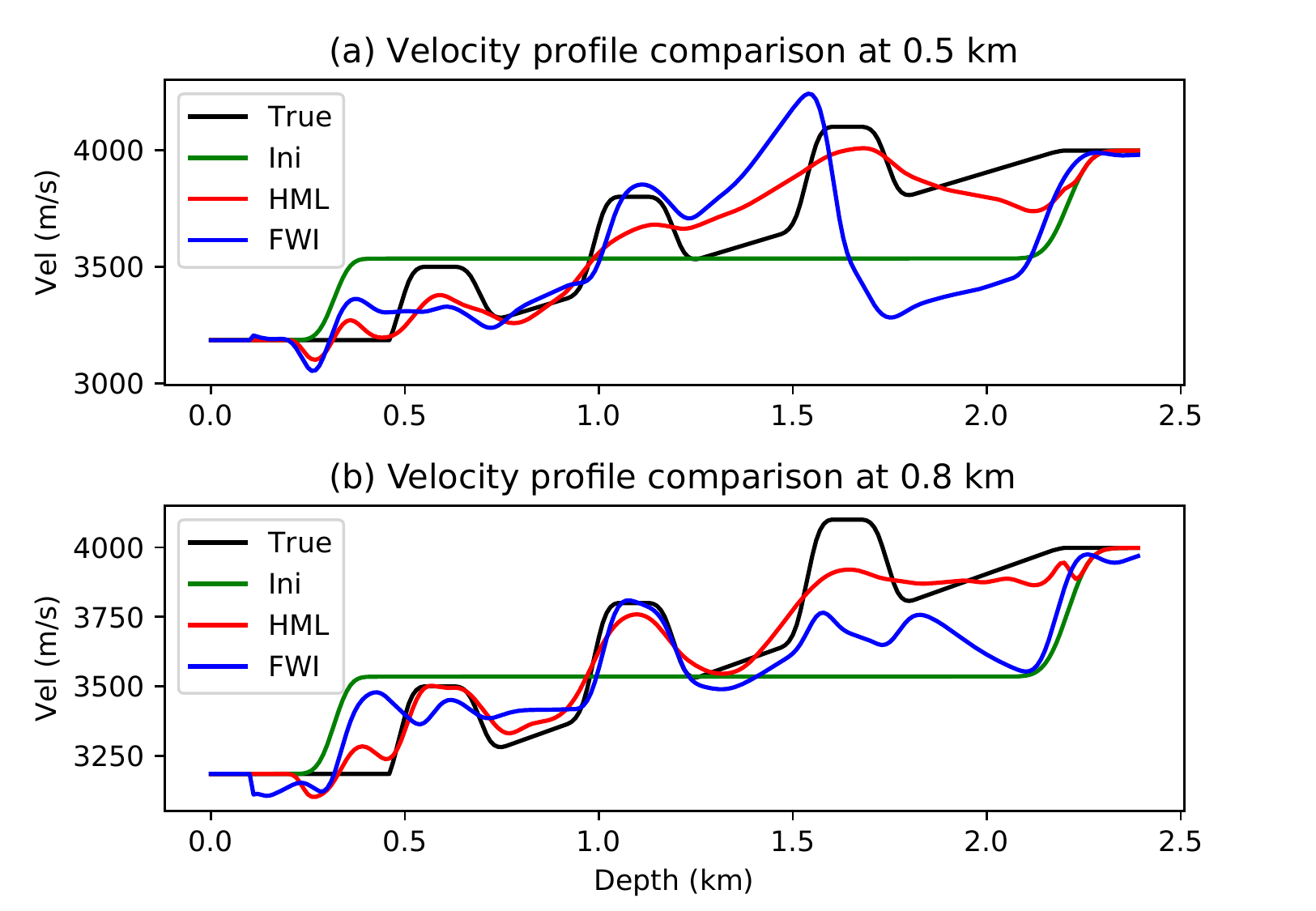}
\caption{The velocity profile comparisons at (a) x= 0.5 km and (b) x = 0.8 km between the true model, initial model, HML and FWI inverted velocity model, which are represent by the black, green, red and blue line, respectively. }
\label{fig:L6}
\end{figure}

We compute the HML misfit $\epsilon$ using equation \ref{eq:misift1} and then uses the AD to automatically calculate the HML ($la=1$) virtual source $\frac{\partial \epsilon}{\partial d}$ which is shown in Figure \ref{fig:L3}a. Compared to the NML virtual source shown in Figure \ref{fig:L3}b which is computed by perturbing the DL feature differences between the observed and predicted data on the predicted shot gather trace by trace, the HML ($la=1$) virtual source is very dissimilar in waveform's shape. The reason is that the latent space dimension is too small to preserve the information of waveform variations. This problem can be solved by using a larger dimensional latent space. However, both the HML ($la=1$) and NML virtual source shows an opposite waveform polarity on the left- and right-hand side of trace $\#70$, which indicates opposite velocity updates on the gradient. The HML ($la=1$) velocity gradient $\frac{\partial \epsilon}{\partial v}$ is estimated by combining the HML ($la=1$) virtual source with the imaging condition. Figure \ref{fig:L4}c shows the first iteration gradient of HML ($la=1$) which is dominated by the low-wavenumber updates. The HML ($la=1$) and FWI inverted model are shown in Figures \ref{fig:L5}a and \ref{fig:L5}b, respectively, where the FWI result suffers severely from the cycle-skipping problem especially at the deep part below z = 1.4 km. Figures \ref{fig:L6}a and \ref{fig:L6}b show the velocity profile comparisons at x = 0.5 km and x = 0.8 km, respectively, between the true, initial, HML ($la=1$) inverted and FWI inverted velocity model, which are represented by the black, green, red and blue line. It clearly shows that HML ($la=1$) has successfully recovered the low wavenumber information of the velocity model. In contrast, the FWI inverted result is far away from the true model.

\begin{figure}[h]
\centering
\includegraphics[width=0.6\columnwidth]{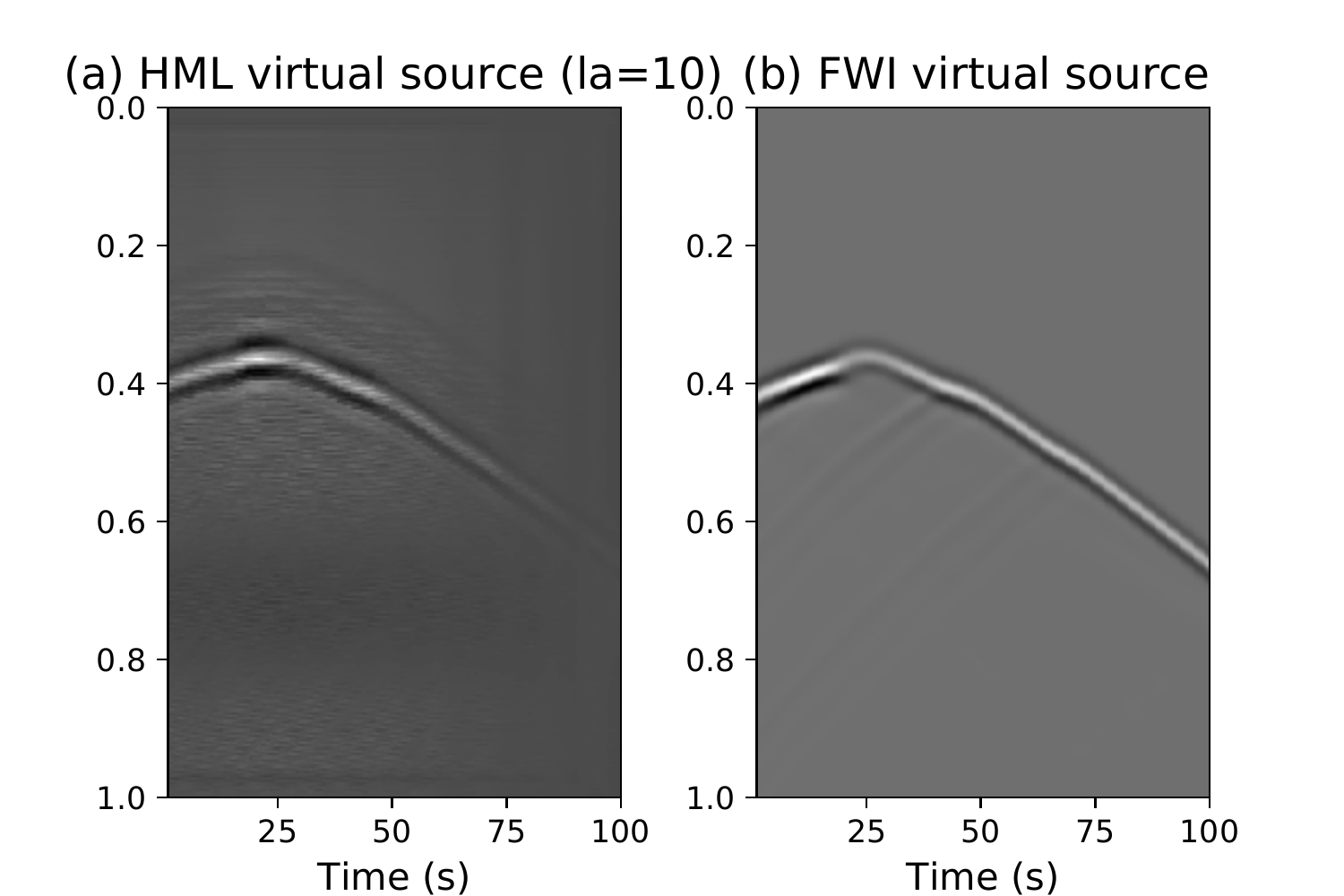}
\caption{The computed (a) HML virtual source with the latent space dimension equals to 10. The (b) FWI virtual source.}
\label{fig:L7}
\end{figure}

\begin{figure}[h]
\centering
\includegraphics[width=1\columnwidth]{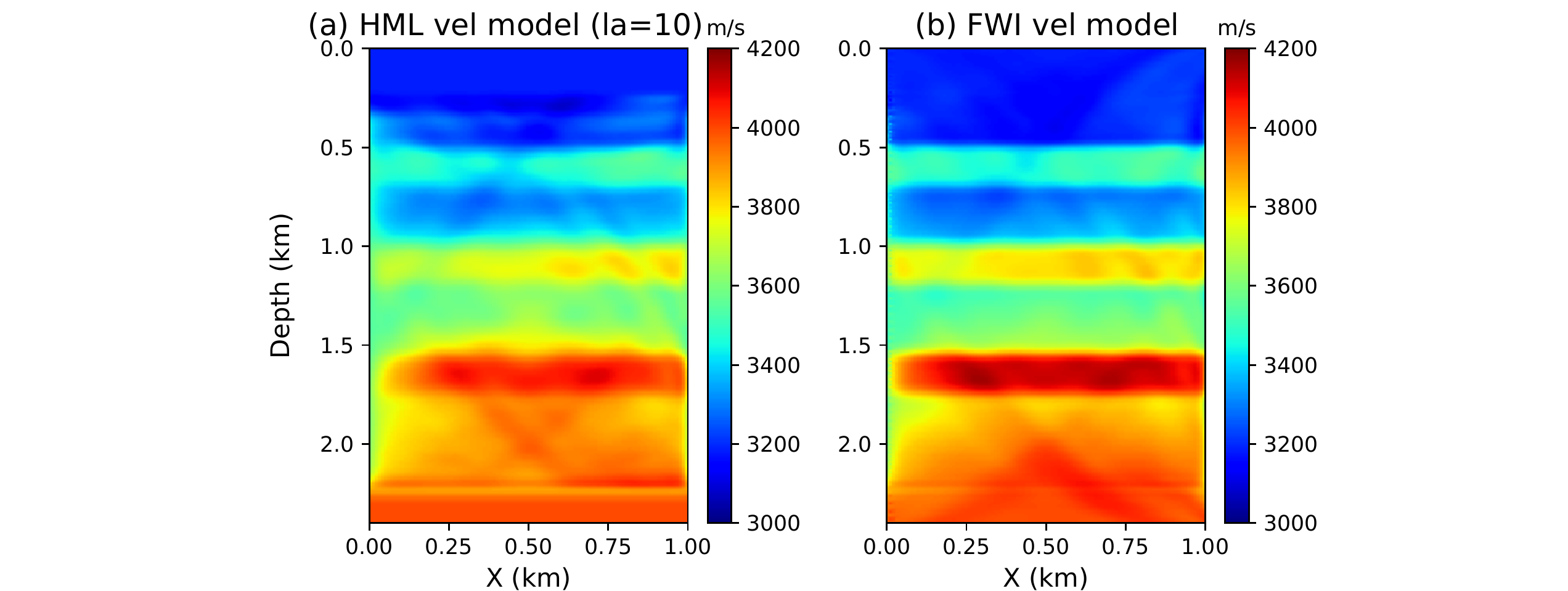}
\caption{The (a) HML ($la=10$) inverted velocity model using the previous HML ($la=1$) inverted result as the initial model. The (b) FWI inverted result which uses (a) as the initial model.}
\label{fig:L8}
\end{figure}

In the next step, we increase the latent space dimension to ten, and re-train the autoencoder using the observed data and the predicted data that generated based on the HML ($la=1$) inverted model. Figure \ref{fig:L7}a shows the computed HML ($la=10$) virtual source which is similar to the FWI virtual source shown in Figure \ref{fig:L7}b. Here, the FWI virtual source is computed by subtracting the predicted data from the observed data. This similarity is because the autoencoder can preserve both the kinematic and dynamic information, such as the traveltime and waveform variations, of the seismic traces by using a larger latent space. Figure \ref{fig:L8}a shows the HML ($la=10$) inverted velocity model where most of the high-wavenumber information has been recovered. To further recover the high-wavenumber details, we use this HML ($la=10$) inverted result as the initial model for FWI inversion. Figure \ref{fig:L8}b shows the FWI inverted result which has the best resolution among all these results.

\subsection{SEAM model}

\begin{figure}[!h]
\centering
\includegraphics[width=1\columnwidth]{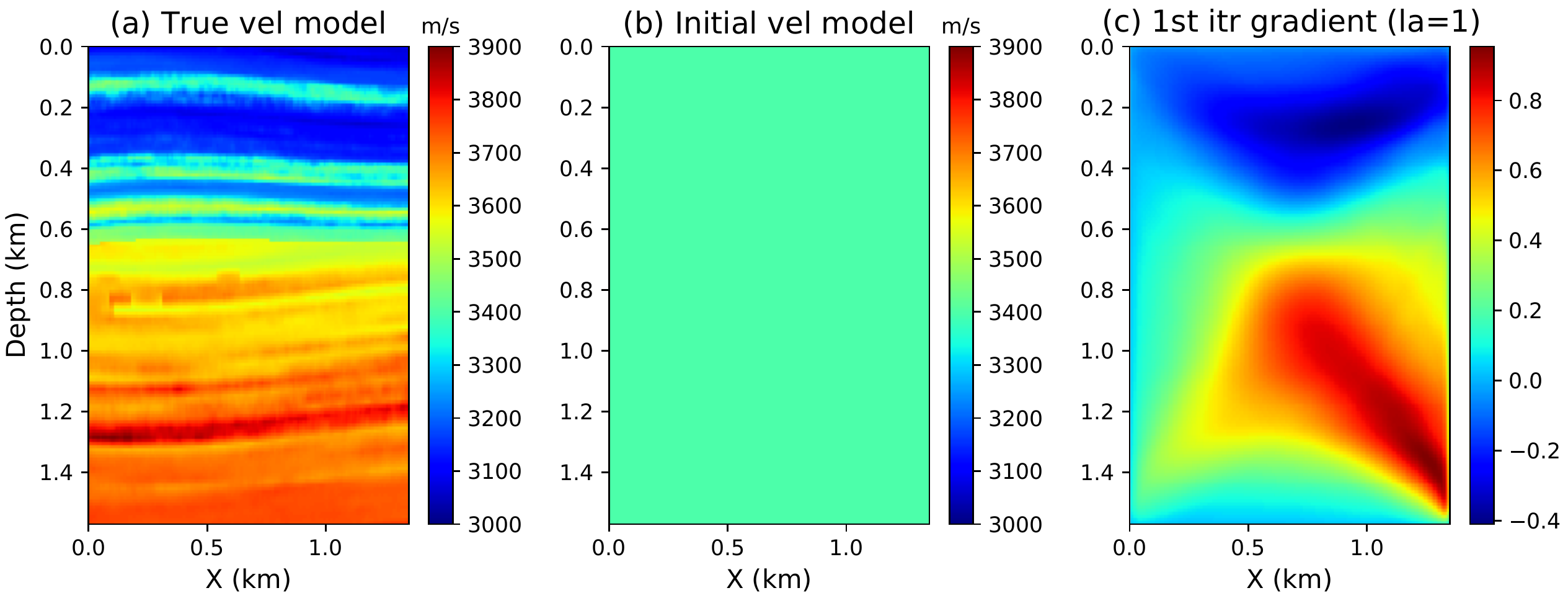}
\caption{The (a) true and (b) initial model. The (c) ist iteration gradient of HML (la=1) inversion.}
\label{fig:S1}
\end{figure}

Data calculated from a portion of the SEAM model with a size of $157 \times 135$ grid point are used to test the HML with the hybrid approach method. Figure \ref{fig:S1}a shows the true model and a homogeneous model is used as the initial model, which is shown in Figure \ref{fig:S1}b. A source well is located at x = 0.01 km and there are 52 shots distributed on the well at an equal spacing of 30 m. Each shot includes 156 receivers which are evenly deployed on the receiver well located at x = 1.35 km. The receiver interval is 10 m and a 20 Hz Ricker wavelet is used as the source. 

\begin{figure}[!h]
\centering
\includegraphics[width=1\columnwidth]{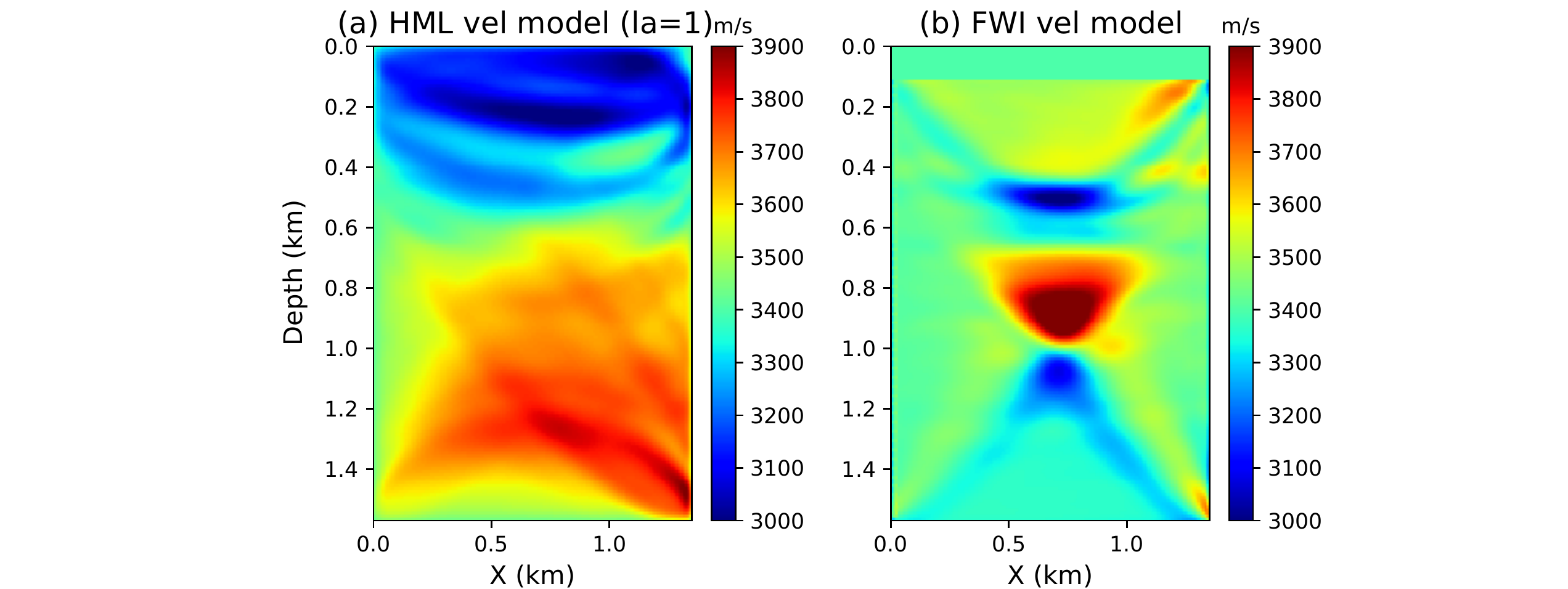}
\caption{The (a) HML (la=1) and (b) FWI inverted result.}
\label{fig:S2}
\end{figure}

\begin{figure}[!h]
\centering
\includegraphics[width=0.7\columnwidth]{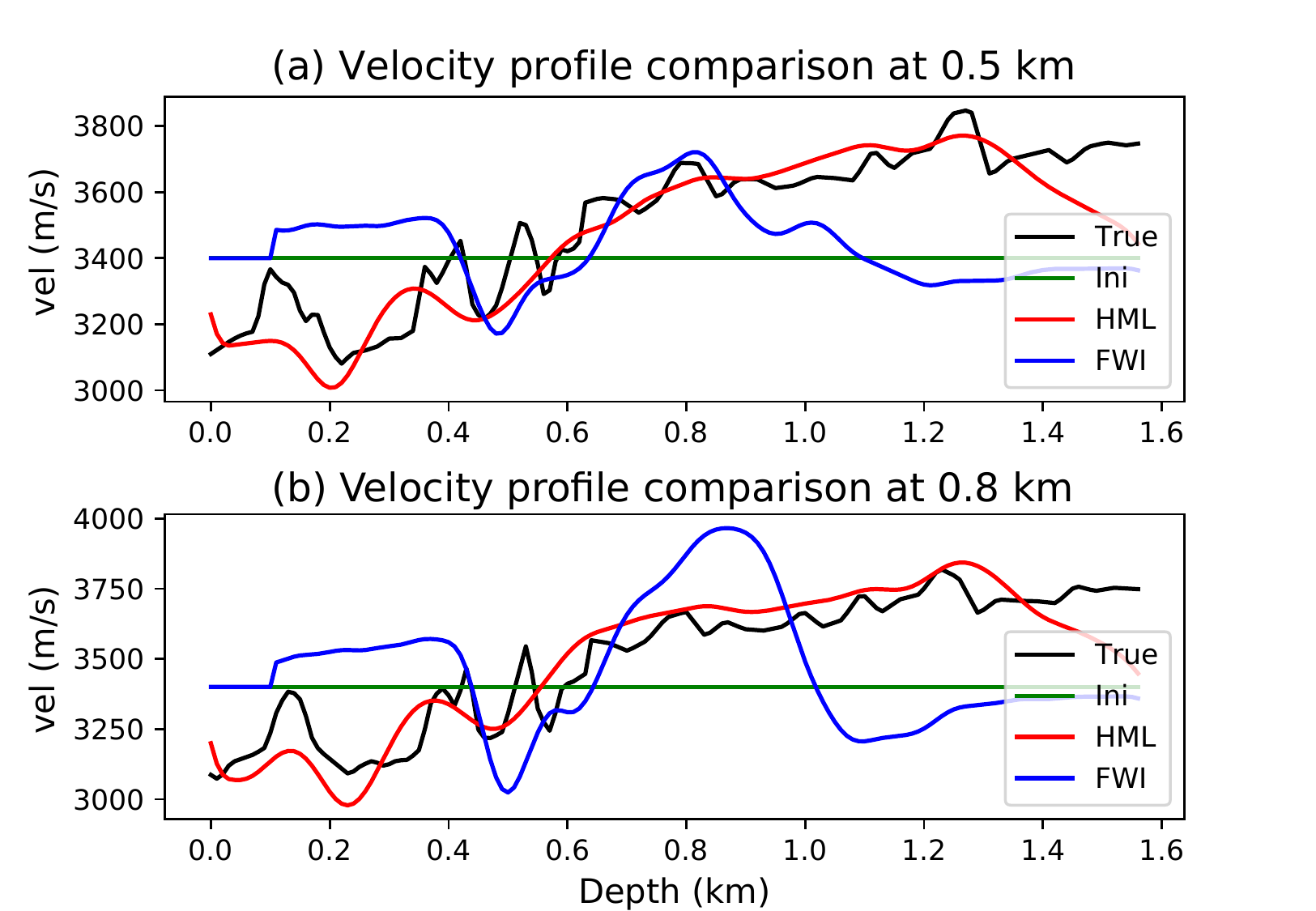}
\caption{The velocity profile comparisons at (a) x= 0.5 km and (b) x = 0.8 km between the true model, initial model, HML and FWI inverted velocity model, which are represent by the black, green, red and blue line, respectively. }
\label{fig:S3}
\end{figure}

Similar to the layered model test, an autoencoder with a one-dimensional latent space is first trained by the observed and predicted seismic traces to learn the one-dimensional DL features that contain the kinematic information of the seismic traces. Once the training is finished, we use HML ($la=1$) with the hybrid approach to invert these DL features for the low-wavenumber information of the subsurface model. Figure \ref{fig:S1}c shows the first iteration gradient of HML ($la=1$) inversion which is dominated by the low-wavenumber updates. Figures \ref{fig:S2}a and \ref{fig:S2}b show the inverted velocity model by HML ($la=1$) and FWI, respectively, where the FWI result suffers severely from the cycle-skipping problem. In comparison, the HML ($la=1$) inverted model has successfully recovered the low-wavenumber information of the subsurface velocity model. This successful recovery can be further proved by the velocity profile comparisons at x = 0.5 km and x = 0.8 km, which are shown in Figure \ref{fig:S3}a and \ref{fig:S3}b, respectively. The black, green, red, and blue curve represents the velocity profile of the true, initial, HML ($la=1$) inverted and FWI inverted velocity model, where the HML ($la=1$) inverted result best matches with the true model. However, the high-wavenumber information is still missing in the HML ($la=1$) inverted result because the latent space dimension is too small to preserve the information of waveform variations of the seismic data. 

\begin{figure}[!h]
\centering
\includegraphics[width=1\columnwidth]{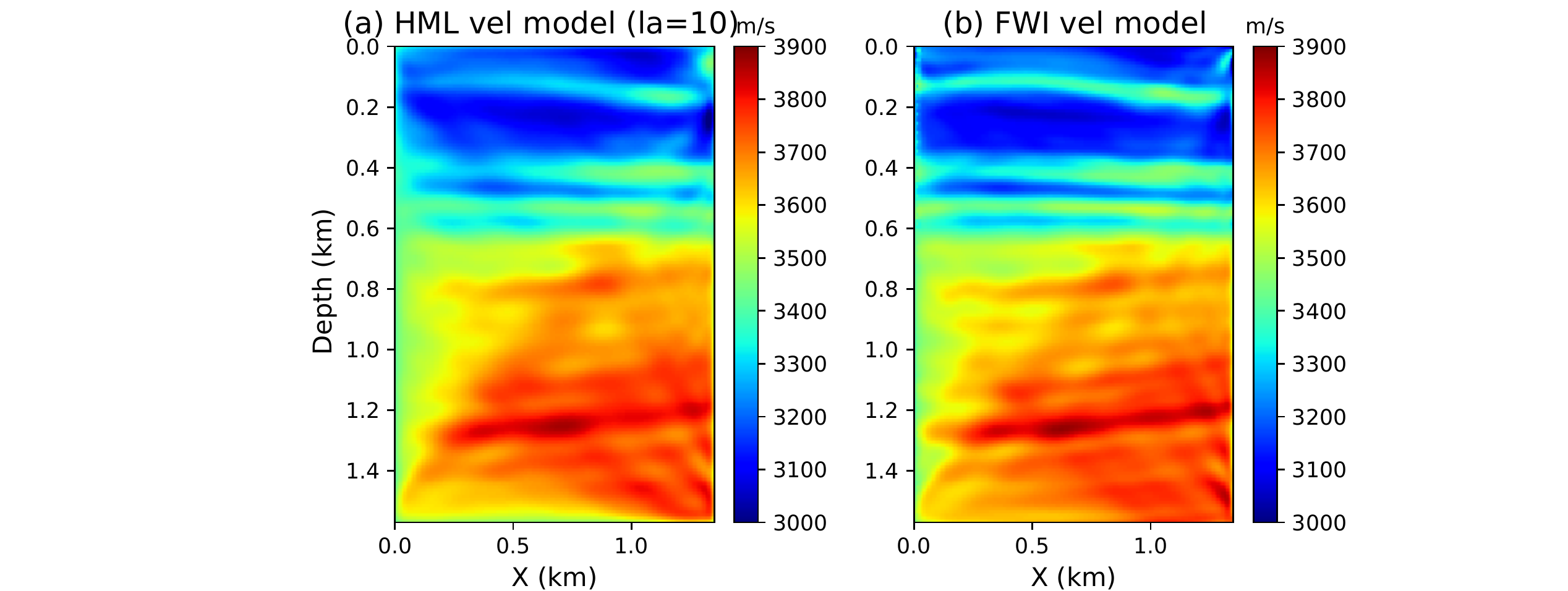}
\caption{The (a) HML ($la=10$) inverted velocity model which uses the HML ($la=1$) inverted result as the initial model. The (b) FWI inverted velocity model which uses the HML ($la=10$) inverted result as the initial model.}
\label{fig:S4}
\end{figure}

Following the multiscale strategy, we increase the latent space dimension to ten and re-train the autoencoder using the observed data and the predicted data that is generated based on the HML ($la=1$) inverted model. We then invert the ten-dimensional DL features for the high-wavenumber information of the velocity model and the HML ($la=10$) inverted result is shown in Figure \ref{fig:S4}a. It shows a obvious resolution increases when compared to the HML ($la=1$) inverted result. Finally, we use FWI to further recover the velocity details and the inverted result is shown in Figure \ref{fig:S4}b, which shows a better resolution at the depth above z = 0.6 km.


\subsection{Gulf of Aqaba field data}
\begin{figure}[!h]
\centering
\includegraphics[width=0.9\columnwidth]{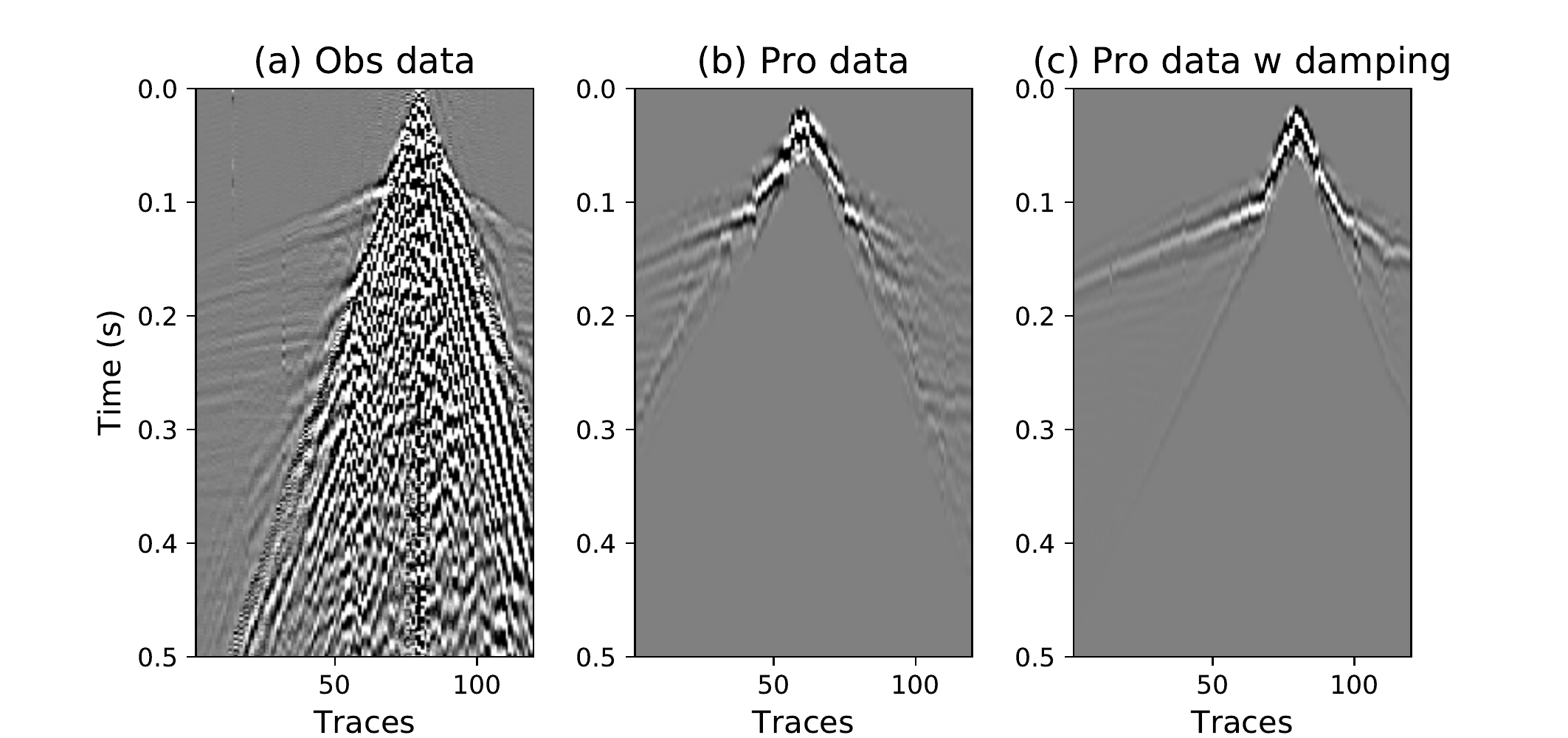}
\caption{An example of a (a) raw and (b) processed shot gather. (c) is the processed shot gather with damping along the time axis.}
\label{fig:A1}
\end{figure}

\begin{figure}[!h]
\centering
\includegraphics[width=1\columnwidth]{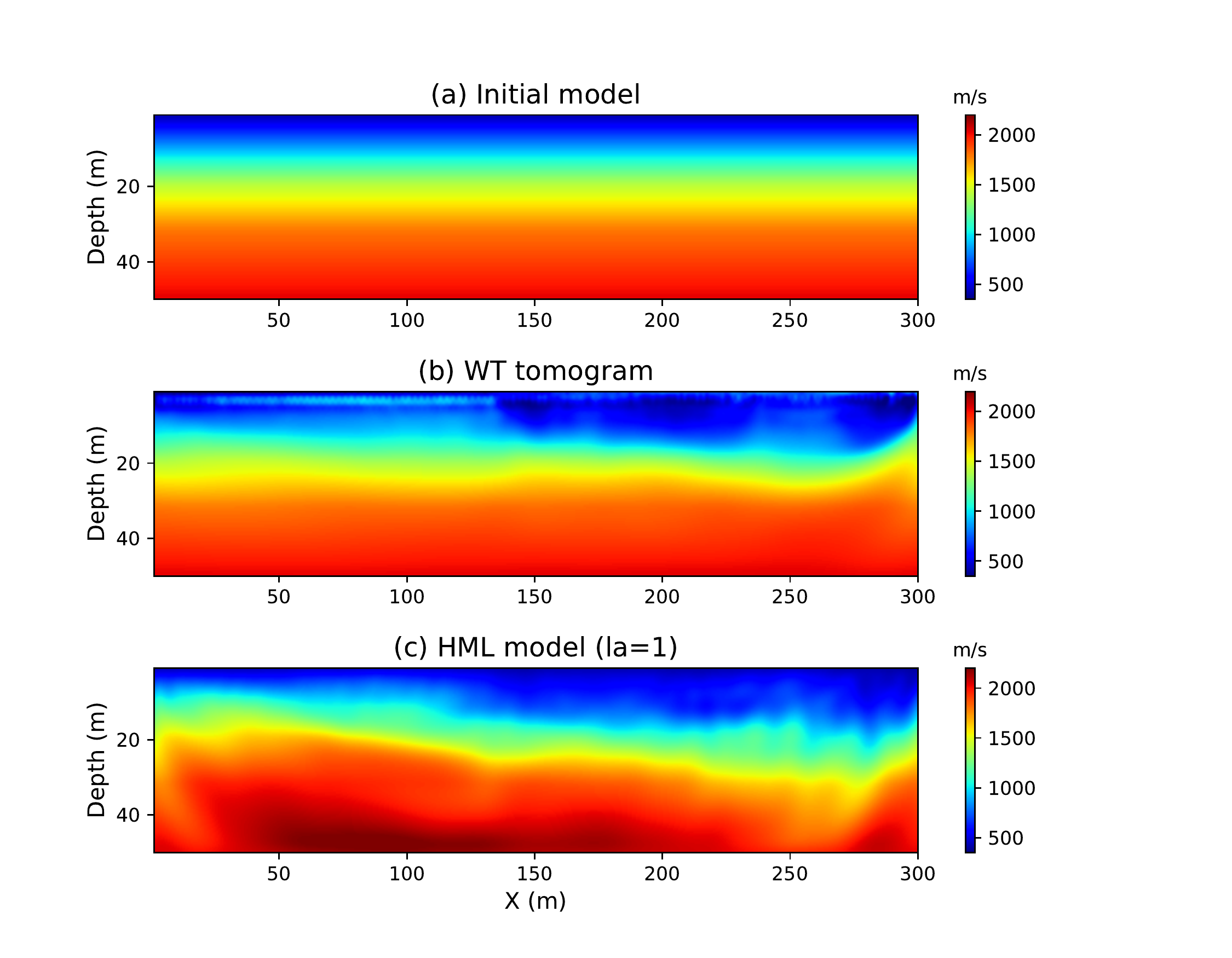}
\caption{The (a) initial model and (b) wave-equation traveltime inversion method inverted model. The (c) HML ($la=1$) inverted velocity model with latent space dimensional equals to one.}
\label{fig:A3}
\end{figure}

The field dataset is collected by a surface seismic survey at an alluvial fan on the Gulf of Aqaba coast in Saudi Arabia. A total of 120 shot gathers were collected at an equal spacing of 2.5 m. Each shot contains 120 traces evenly distributed on the seismic survey with a receiver interval of 2.5 m. Data were recorded using a 1 ms sampling rate for total recording time of 0.5 s. A 200 lb weight drop was used as the source, with 10 to 15 stacks at each shot location \citep{hanafy2014imaging}. An example of a raw shot gather is shown in Figure \ref{fig:A1}a which includes very strong surface wave energy and weak refraction events. We first remove the surface waves because we only consider inverting the P waves in this paper. We then bandpass the data to the peak-frequency of 40 Hz. A processed shot gather is shown in Figure \ref{fig:A1}b, where only the refractions event remains. We further apply an amplitude damping on the time axis to highlight the early arrivals and attenuate the later arrivals. One example of the processed + damping shot gather is shown in Figure \ref{fig:A1}c, where the early arrivals has been highlighted. A linear increasing model shown in Figure \ref{fig:A3}a is used as the initial model. 

\begin{figure}[!h]
\centering
\includegraphics[width=1\columnwidth]{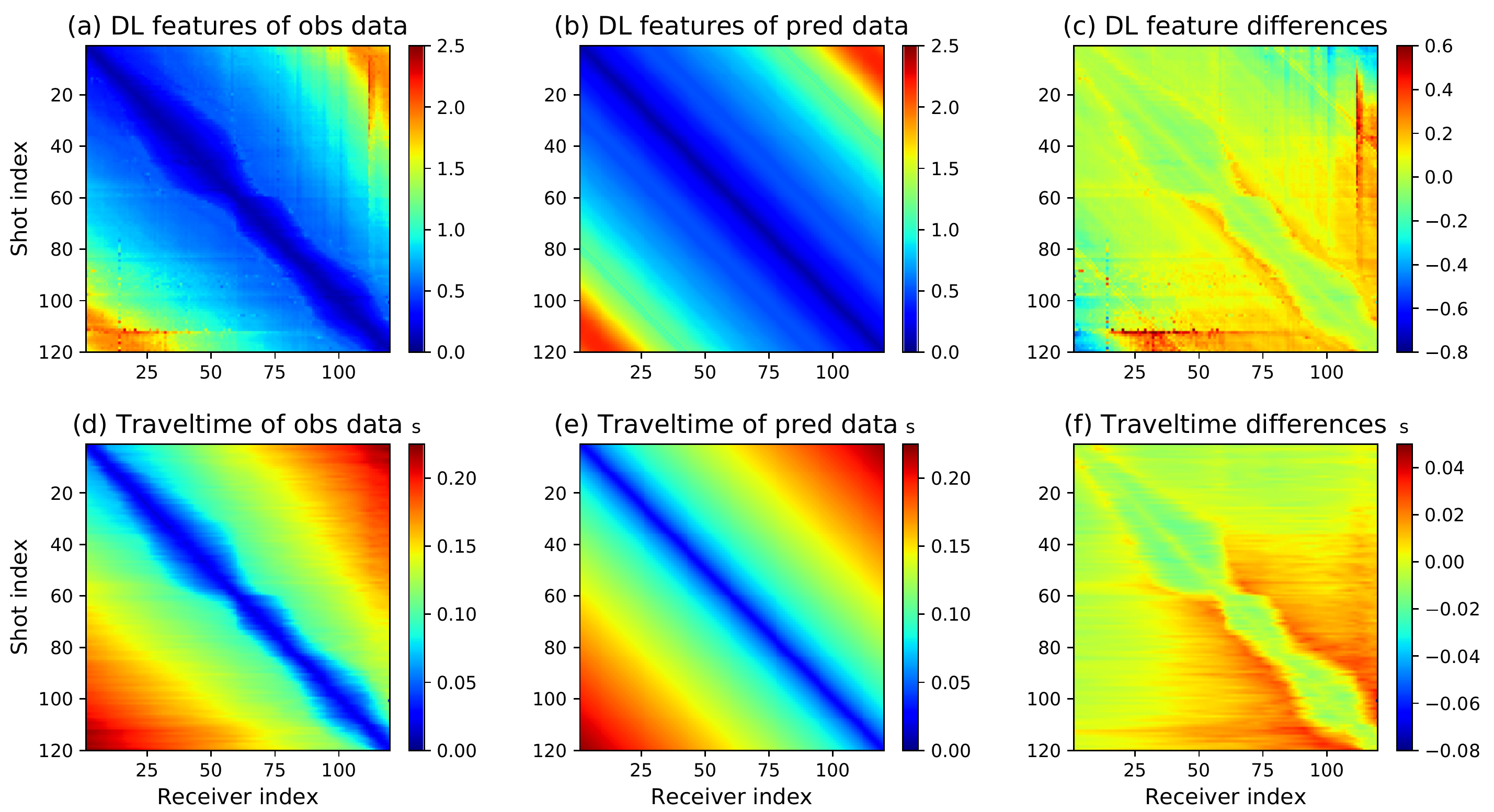}
\caption{The DL features of the (a) observed and (b) predicted data, where the predicted data is generated based on the initial model. The (c) DL differences between the observed and predicted data. The traveltime of the (d) observed and (e) predicted data. (f) Their traveltime differences.}
\label{fig:A2}
\end{figure}

\begin{figure}[!h]
\centering
\includegraphics[width=1\columnwidth]{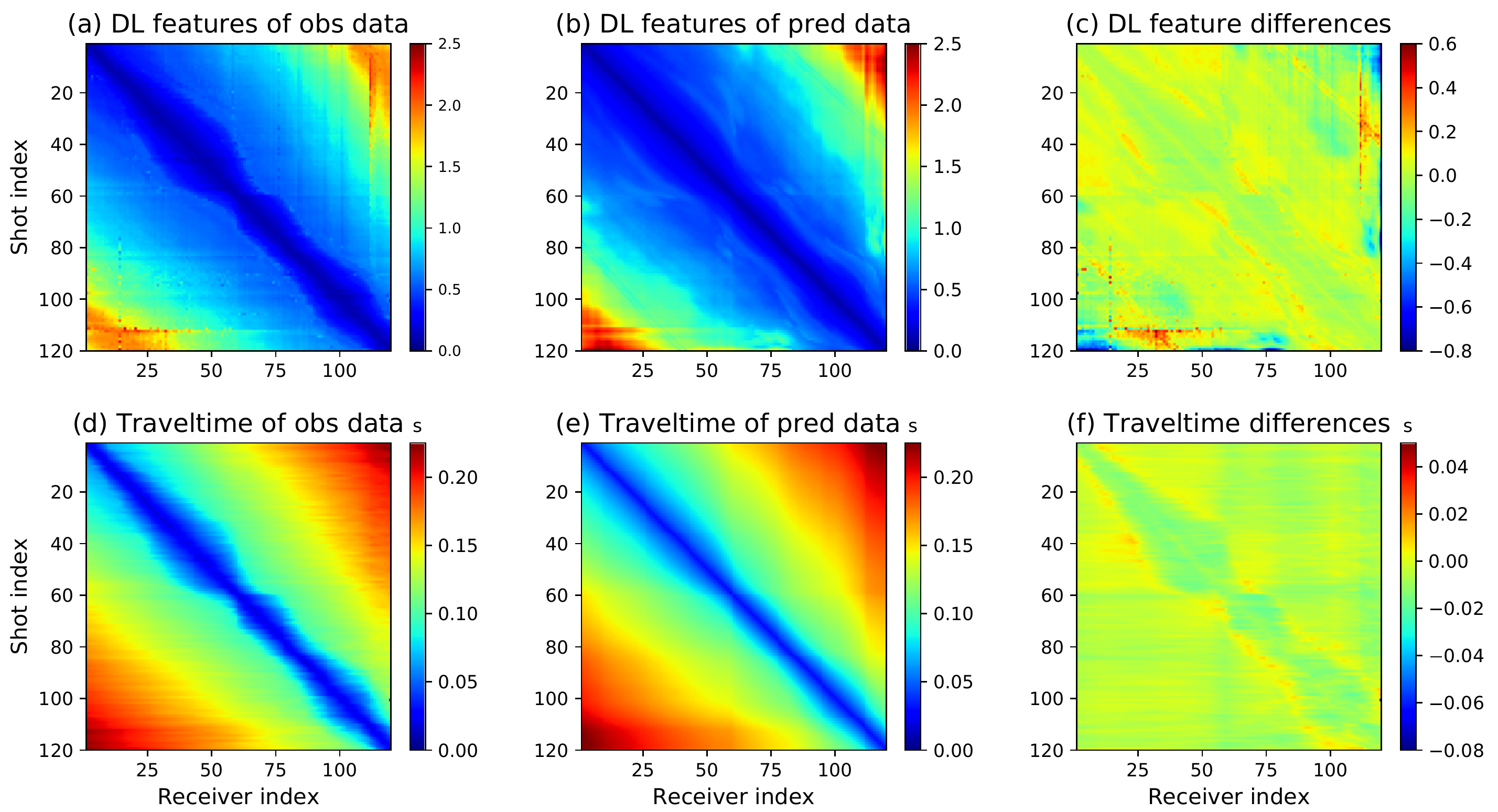}
\caption{The DL features of the (a) observed and (b) predicted data, where the predicted data is generated based on the HML (la=1) inverted result. The (c) DL differences between the observed and predicted data. The traveltime of the (d) observed and (e) predicted data. (f) Their traveltime differences.}
\label{fig:A4}
\end{figure}

According to the multiscale inversion strategy of HML, we first invert the low-dimensional DL features for the background velocity model. A CAE with a single-dimensional latent space is first trained using the seismic traces from the processed + damping shot gathers. The well-trained CAE can effectively compress the $nt \times 1$ seismic traces to the $1\times1$ DL features. To make sure that the compressed DL features mainly contains the kinematic information of the seismic traces, we compare the DL features with the traveltimes. Figure \ref{fig:A2}a shows the compressed DL feature map of the observed data, where the vertical and horizontal axis indicates the shot and receiver index. Each pixel in this figure represents the compressed $1 \times 1$ DL feature value of the seismic trace for a certain shot-receiver pair. Figure \ref{fig:A2}d shows the traveltime map of the observed data, which shows a similar pattern to Figure \ref{fig:A2}a. Figures \ref{fig:A2}b and \ref{fig:A2}e show the DL feature and traveltime map of the predicted data, which also shows a similar pattern. The most obvious similarity between the DL features and traveltimes can be seen in their difference map shown in Figures \ref{fig:A3}c and \ref{fig:A3}f, respectively. Both the DL feature and traveltime differences show that the major difference between the observed and predicted data is within the area between shot $\#40$ to $\#120$ and receiver $\#50$ to $\#120$. Therefore, we can conclude that the compressed one-dimensional DL features do contain the kinematic information of the seismic traces, which is similar to the traveltime. Figure \ref{fig:A3}b shows the inverted velocity model using the wave-equation traveltime (WT) inversion, which reveals a dipping interface between the upper low-velocity layer and the bedrock. This dipping feature is because that the mountain and the sea are located on the left- and right-hand side of the seismic survey, respectively. This dipping can be seen more clearly in the HML ($la=1$) inverted result which is shown in Figure \ref{fig:A3}c. After inversion, we generate a new set of predicted shot gathers based on the HML ($la=1$) inverted velocity model. The DL feature and traveltime maps of the new predicted data are shown in Figures \ref{fig:A4}b and \ref{fig:A4}e, which is similar to their corresponding observed maps that are shown in Figure \ref{fig:A4}a and \ref{fig:A4}d. Their differences are shown in Figures \ref{fig:A4}c and \ref{fig:A4}f which are much smaller compared to the initial differences shown in Figures \ref{fig:A3}c and \ref{fig:A3}f. The reduced differences indicate that the HML ($la=1$) inverted velocity model is more close to the true model compared to the initial model. 

\begin{figure}[!h]
\centering
\includegraphics[width=1\columnwidth]{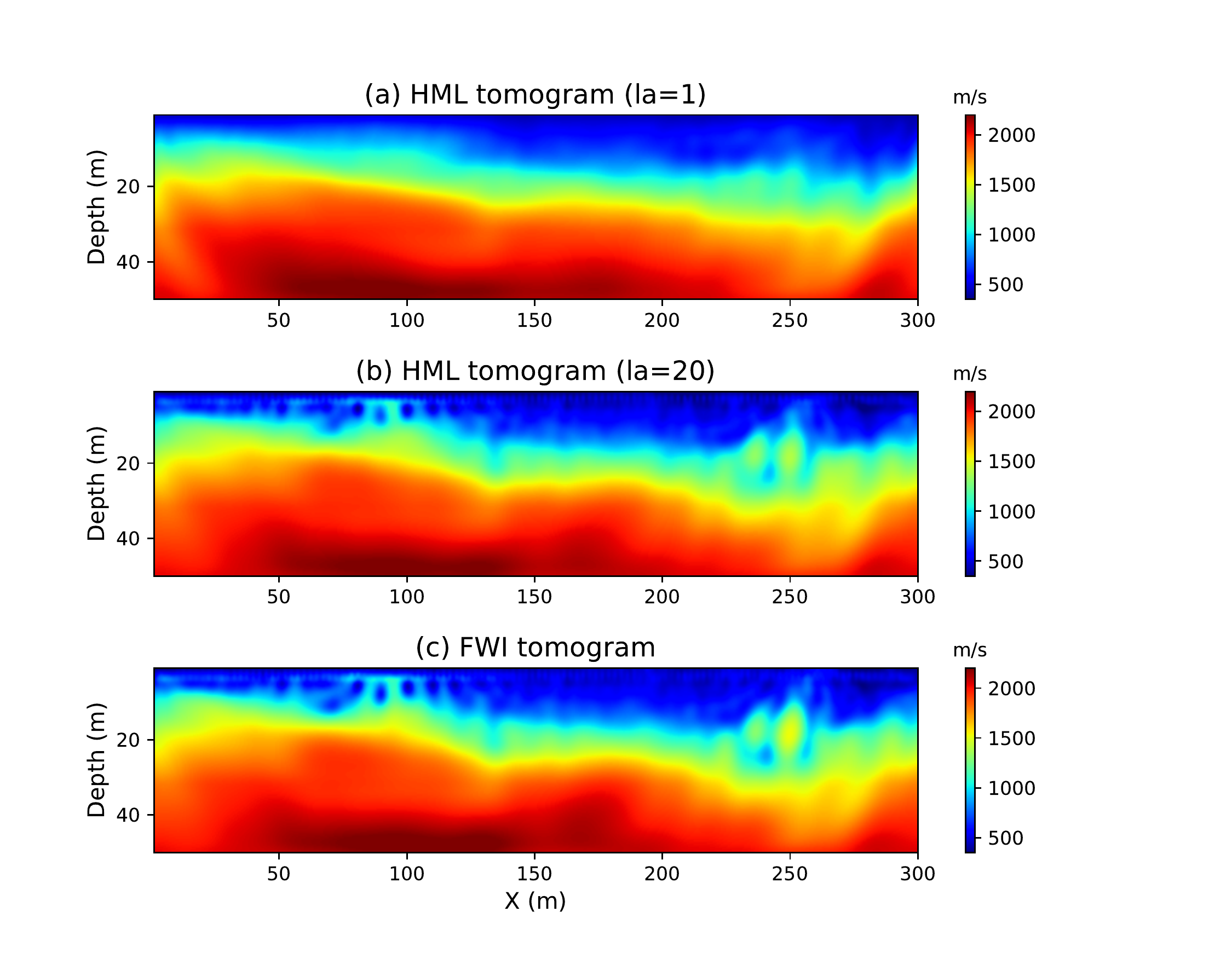}
\caption{(a) The HML ($la=1$) inverted velocity model. (b) The HML ($la=20$) inverted velocity model which uses (a) as the initial model. (c) The FWI inverted result which uses (b) as the initial model.} 
\label{fig:A5}
\end{figure}

\begin{figure}[!h]
\centering
\includegraphics[width=1\columnwidth]{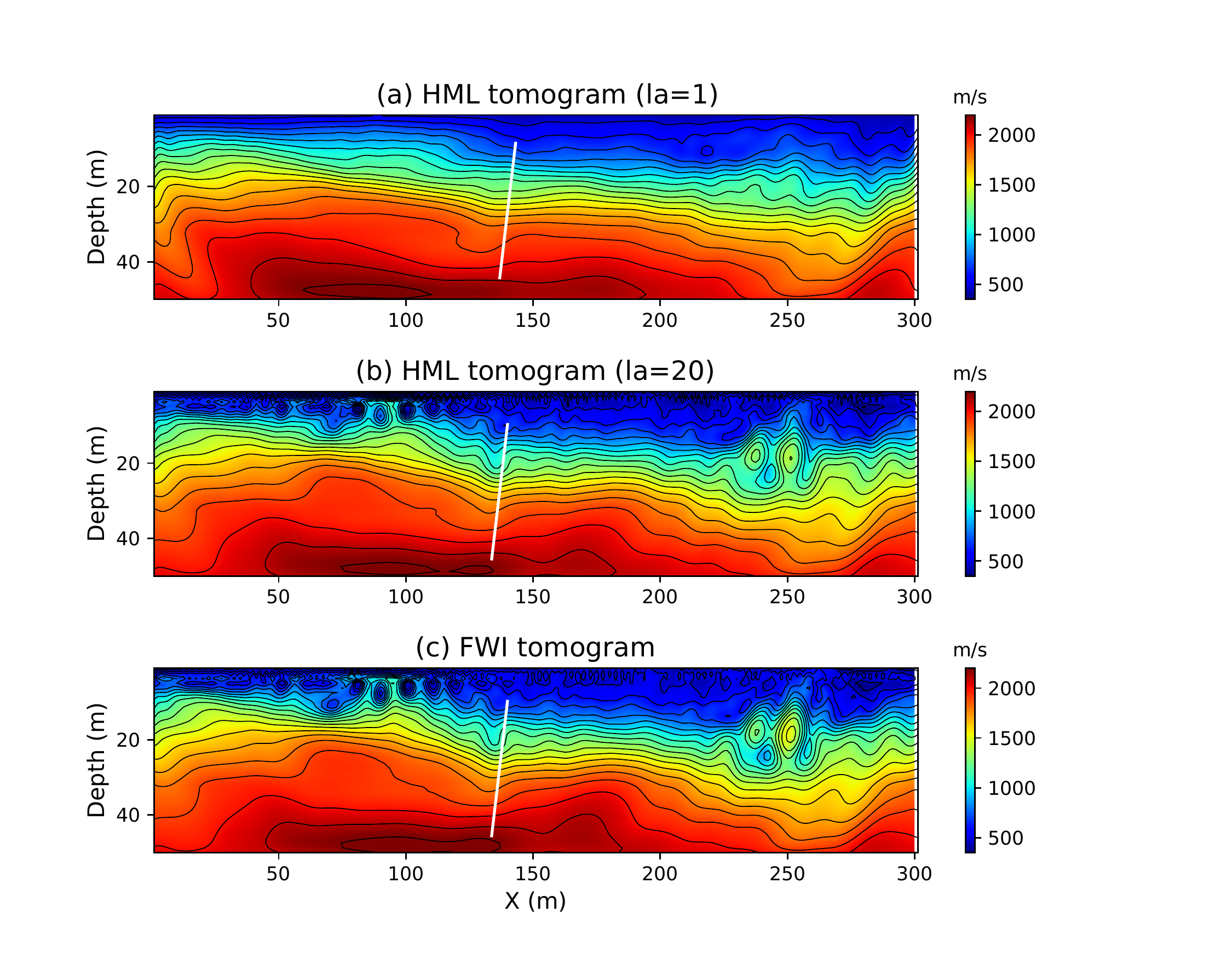}
\caption{(a) The HML ($la=1$) inverted velocity model with overlaped contour lines. (b) The HML ($la=20$) inverted velocity model with overlaped contour lines which uses (a) as the initial model. (c) The FWI inverted result with overlaped contour lines which uses (b) as the initial model. The white line indicates the fault.}
\label{fig:A6}
\end{figure}

In the next step, we use the HML ($la=1$) inverted velocity model as the initial model and start to recover the high-wavenumber information of the subsurface model. We increase the latent space dimension to twenty and re-train the autoencoder using the seismic traces from the processed shot gathers. The reason we use the processed rather than the processed plus damping shot gathers for training is that the twenty-dimensional latent space is capable of preserving the kinematic and dynamic information for both the early and later P wave events. The HML ($la=20$) inverted velocity model is shown in Figure \ref{fig:A5}b which reveals more high-resolution details compared to the HML ($la=1$) inverted result. There are some low- and high-velocity anomalies appear at the region between x = 80 m to x = 130 m and x = 230 m to x = 280 m, respectively. There also shows a velocity discontinuity at x = 140. This discontinuity could be caused by an active fault which has been identified by \cite{hanafy2014imaging}. Figure \ref{fig:A5}c shows the FWI inverted model which uses the HML ($la=20$) inverted result as the initial model. The FWI method slightly increased the velocity resolution which means the HML ($la=20$) inverted result is already good enough. Figures \ref{fig:A6}a, \ref{fig:A6}b and \ref{fig:A6}c show the HML ($la=1$), HML ($la=20$) and FWI inverted velocity model overlaped with their contour lines. The contour lines around x = 140 m in Figures \ref{fig:A6}b and \ref{fig:A6}c point downward which further highlight the velocity discontinuity in this region. We mark the possible fault using the white line on these figures.

\section{Conclusion}
We present a seismic inversion method that inverts the deep learning (DL) features for the subsurface velocity model. The DL feature is a low-dimensional representation of the high-dimensional seismic data, which is automatically generated by a convolutional autoencoder (CAE) and preserved in the latent space. When the latent space dimension is small, the DL feature mainly contains the kinematic information, such as the traveltime, of the input seismic data. However, both the kinematic and dynamic information, such as the traveltime and waveform variations, can be preserved in the DL features by using a larger latent space. Therefore, we propose a multiscale inversion approach which starts with inverting the low-dimensional DL features for the low-wavenumber information of the subsurface model. Then recover its high-wavenumber details through inverting the high-dimensional DL features. 

Because there are no governing equations that contain both the velocity and DL feature term in the same equation, therefore we use the automatic differentiation (AD) to numerically connect these two terms together. In another word, we use the AD to numerically connect the CAE network with the wave-equation inversion. One can replace the CAE network with any type of deep learning architecture and connected with any type of Newton equations by using the AD to solve various problems. We denote this hybrid connection through the AD as hybrid machine learning (HML). This method would be appreciated by geophysical novices because the AD replaces the complex math derivation with a black box so anyone can do HML without having a deep background in geophysics. However, one concern of the HML method is it computational costs because it is expensive to use the AD to solve the wave-equation inversion. Therefore we also propose a hybrid implementation approach which makes HML has the same level of computational efficiency compared to the conventional wave-equation method, such as full waveform inversion (FWI). This hybrid implementation approach brings HML the potential of solving a very large scale inversion problem. 



\section{Acknowledgement}
This research was fully funded by the Deep Earth Imaging Future Science Platform, CSIRO. 


\bibliographystyle{seg}  
\bibliography{paper}

\listoffigures

\end{document}